# FLARE FOOTPOINT REGIONS AND A SURGE OBSERVED BY THE *HINODE*/EUV IMAGING SPECTROMETER (EIS), *RHESSI*, AND *SDO*/AIA

G. A. Doschek

*Space Science Division, Naval Research Laboratory, Washington, DC 20375, USA*

H. P. Warren

*Space Science Division, Naval Research Laboratory, Washington, DC 20375, USA*

B. R. Dennis

*Solar Physics Laboratory, Heliophysics Science Division, NASA Goddard Space Flight Center, Greenbelt, MD 20771, USA*

J. W. Reep

*Department of Physics and Astronomy, Rice University, Houston, TX 77005, USA*

A. Caspi

*Southwest Research Institute, Boulder, CO 80302, USA*

## ABSTRACT

The Extreme-ultraviolet Imaging Spectrometer (EIS) on the *Hinode* spacecraft observed flare footpoint regions coincident with a surge for a M3.7 flare observed on 25 September 2011 at N12 E33 in active region 11302. The flare was observed in spectral lines of O VI, Fe X, Fe XII, Fe XIV, Fe XV, Fe XVI, Fe XVII, Fe XXIII and Fe XXIV. The EIS observations were made coincident with hard X-ray bursts observed by the *Reuven Ramaty High Energy Solar Spectroscopic Imager* (*RHESSI*). Overlays of the *RHESSI* images on the EIS raster images at different wavelengths show a spatial coincidence of features in the *RHESSI* images with the EIS upflow and downflow regions, as well as loop-top or near-loop-top regions. A complex array of phenomena was observed including multiple evaporation regions and the surge, which was also observed by the *Solar Dynamics Observatory* (*SDO*)/Atmospheric Imaging Assembly (AIA) telescopes. The slit of the EIS spectrometer covered several flare footpoint regions from which evaporative upflows in Fe XXIII and Fe XXIV lines were observed with Doppler speeds greater than 500 km s$^{-1}$. For ions such as Fe XV both evaporative outflows ($\sim 200$ km s$^{-1}$) and downflows ($\sim 30 - 50$ km s$^{-1}$) were observed. Non-thermal motions from 120 to 300 km s$^{-1}$ were measured in flare lines. In the surge, Doppler speeds are found from about 0 to over 250 km s$^{-1}$ in lines from ions such as Fe XIV. The non-thermal motions could be due to multiple sources slightly Doppler-shifted from each other or turbulence in the evaporating plasma. We estimate the energetics of the hard X-ray burst and obtain a total flare energy in accelerated electrons of $\geq 7$ x $10^{28}$ ergs. This is a lower limit because only an upper limit can be determined for the low energy cutoff to the electron spectrum. We find that detailed modeling of this event would require a multi-threaded model due to its complexity.

*Subject headings:* Sun: flares, Sun: activity, Sun: UV radiation, Sun: corona, Sun: X-rays, gamma rays

## 1. INTRODUCTION

The advent of imaging spectroscopy has led to much work towards understanding chromospheric evaporation in flares in the context of the so-called Standard Flare Model (e.g., Shibata 1996), where energy is released due to magnetic reconnection in the corona or as a by-product of this reconnection. The most recent work has utilized spectra from the Extreme-ultraviolet Imaging Spectrometer (EIS) (Culhane et al. 2007) on the *Hinode* spacecraft (e.g., Milligan & Dennis 2009; Watanabe et al. 2010; Hara et al. 2011; Brosius 2013; Young et al. 2013; Imada et al. 2013; Doschek et al. 2013). One of the goals of this work is to test evaporative models where chromospheric heating and evaporation are produced at the footpoints of reconnecting flare loops by the impact of energetic particles accelerated in or near the reconnection region. The X-ray and/or EUV emission from such particles can be detected by the *Reuven Ramaty High Energy Solar Spectroscopic Imager* (*RHESSI*) spacecraft and also by spot-like enhancements in *Solar Dynamics Observatory* (*SDO*)/Atmospheric Imaging Assembly (AIA) images such as the 1700 Å continuum. *RHESSI* is described by Lin et al. (2002) and *SDO* and AIA are described by Pesnell et al. (2012) and Lemen et al. (2012), respectively.

Numerical models predict a dynamical chromospheric response to the heating that is a function of temperature and energy input. Because of the limited field-of-view of a slit spectrometer, obtaining flare footpoint spectra and associating the spectra with data from other spacecraft have been challenging and dependent on fortuitous circumstances. A flare coincident with a surge that occurred on 25 September 2011 near 15:31 UT provides such circumstances. The slit of the EIS spectrometer was being stepped across an active region and fortuitously crossed *RHESSI*-imaged flare footpoint regions at the time of a group of hard X-ray bursts. Thus, this flare provides an excellent opportunity to observe the chromospheric response to energetic electron input.

In this paper we discuss the EIS spectra, the *RHESSI* hard X-ray data, and context data from AIA on *SDO*. For flare footpoint regions and the surge we determine upflow speeds from spectral line profiles of coronal ions ($\sim$ 1-3 MK) and lines from flare ions formed at multi-million degree temperatures ($\sim$ 10-15 MK). Henceforth, the term "coronal ions" refers to ions formed near ambient coronal temperatures. We determine electron densities from an Fe xiv line ratio formed near 2 MK. We determine the energy in non-thermal electrons from *RHESSI* hard X-ray observations. It will become clear that the flare contains many flux tubes, and therefore we do not attempt a detailed numerical model of this event. However, in the summary we discuss briefly the predictions of a single loop simulation with some of the parameters estimated from our results using the HYDRAD code (e.g., Bradshaw & Klimchuk 2011; Reep et al. 2013) for a single flux tube filled with a multi-fluid plasma.

We have found that the region of hard X-ray energy deposition is complex, featuring multiple footpoints that produce either evaporation or contribute to the extensive surge that is observed. The hard X-ray burst is similarly complex, featuring multiple intensity peaks. Doppler speeds in excess of 600 km s$^{-1}$ are observed in the 10 MK or higher temperature line of Fe xxiv at 192 Å. From the position of the flare on the solar disk, this translates to a radial outflow speed greater than 800 km s$^{-1}$. The surge shows plasma at coronal temperatures (about 2 MK) that flows outward at different speeds at different slit locations. The electron density near 2 MK is about $2 \times 10^{10}$ cm$^{-3}$ over much of the flare region. The data support a multi-magnetic-thread model that is consistent with multi-magnetic-island acceleration of electrons in the flare reconnection region.

A description of the flare and results are presented in Section 2. A summary and discussion is given in Section 3. Appendix A presents iron-line spectra separated by single pixels along the EIS slit.

## 2. OBSERVATIONAL RESULTS

### 2.1. EIS Results and Relationships to *RHESSI*, and *SDO* Data

The EIS imaging spectrometer on *Hinode* observes two narrow wavelength bands between about 170-213 Å and 250-290 Å. A multi-layer coated telescope feeds a spectrometer through one of four possible slit/slot apertures oriented in the north-south direction. A 2″ wide slit was used for the flare observations to be described. There are a number of EIS flare studies, i.e., software programs that select spectral lines, spectral window sizes, exposure times, and type of raster, for observations. The flare we observed was obtained with the study: hh_flare_raster_v2. In this study, the slit is stepped across the flare region in 5″ increments, west to east, and an 8 s exposure is recorded at each step position. A total of 36 exposures were made over a raster area of about 180″ x 160″. The spectra include lines of O VI, Fe X, Fe XII (192, 195 Å), Fe XIV (264, 274 Å), Fe XV, Fe XVI, Fe XVII, Fe XXIII, and Fe XXIV (192, 255 Å). The raster was performed repeatedly, but only one of the rasters is relevant for this work. The EIS flare spectra have been corrected using standard EIS software for dark current, the CCD pedestal, orbital temperature variations, warm pixels, and slit tilt.

An image of EIS raster spectra is made by integrating the spectra over wavelength to obtain the total intensities in the spectral lines. The spectra are then stacked from west to east and displayed with IDL routines such as PLOT_IMAGE and TVSCL. For cases where the X-position changes by more than 1″, such as for the data in this paper, the X-axis in the plots is still spatially scaled to 1″ by using the IDL SCALE keyword in PLOT_IMAGE. The result is that images from such rasters are composed of vertical bars that are clearly seen. However, the file numbering system for the images is from east to west, i.e., the spectrum with file number 0 is the last spectrum obtained and is at the eastern border of the image. These comments should help the reader understand the EIS images that are shown below.

Important dynamical quantities obtained from the EIS spectra are Doppler shift speeds and non-thermal motions. Non-thermal motions are derived from the profiles of the spectral lines. We have fit data using the IDL GAUSSFIT routine. The full width at half maximum (FWHM) of a spectral line is given by

$$FWHM = 1.665 \, \frac{\lambda}{c} \, \sqrt{\frac{2kT}{M} + V^2 + W_I^2} \quad , \tag{1}$$

where $\lambda$ is the wavelength (in Å in this paper), $c$ is the speed of light, $k$ is the Boltzmann constant, $T$ is the electron temperature (degrees Kelvin), $M$ is the ion mass, $V$ is the non-thermal motion, and $W_I$ is the instrumental width.

An uncomplicated symmetric spectral line is broadened by three components which are all assumed Gaussian with the same rest wavelengths, i.e., instrumental broadening and two Doppler broadening mechanisms. EIS lines are considerably broadened by the instrumental width, $W_I$. For the 2″ slit this width varies across the CCD in the north-south direction. The FWHM instrumental width at each pixel location has been removed using EIS Software Note No. 7 prepared by Peter Young. (The FWHM is $1.665 \times W_I$ in equation (1)). In addition to instrumental broadening, lines are broadened by thermal

Doppler motions at the temperature $T$ where they are formed in ionization equilibrium. A third broadening mechanism is Doppler broadening due to non-thermal motions $V$. These non-thermal motions might be due to turbulence, but they might also be due to unresolved spectral lines with small Doppler shifts relative to each other. Finally, the temperature in equation (1) is really the ion temperature. We assume that at flare electron densities the ion and electron temperatures are equal because for a typical electron density of about $3 \times 10^{10}$ cm$^{-3}$ and a temperature of about 10 MK, the energy equipartition time between electrons and iron ions is about 1 s (see Spitzer, The Physics of Fully Ionized Gases). Removal of the instrumental and thermal Doppler broadening yields the non-thermal velocities $V$ using equation (1).

We note here that in this paper the word "non-thermal" is used to describe two distinct physical quantities. One quantity is the energy in electrons that are not described by a thermal Maxwellian velocity distribution. These are the non-thermal electrons measured by *RHESSI*. The other quantity is the motion of plasma that produces a Doppler line width in spectral lines that is in excess of the expected thermal width. These motions are generally interpreted as due to a non-thermal process such as turbulence and therefore are called non-thermal motions.

Rest wavelengths of the lines discussed in this paper have been determined by making detailed corrections to the nominal EIS wavelengths given by the EIS data correction program. However, the wavelengths given by the EIS correction program are not necessarily the rest wavelengths. An additional first order correction has been made that substantially improves these wavelengths. Nevertheless, the improved wavelengths are still not precisely what we feel are the best rest wavelengths. For large velocities more than about 100 km s$^{-1}$ these differences are not important. But the downflow speeds we discuss are on the order of 50 km s$^{-1}$ and are more sensitive to these differences. To correct for this, we determine a local rest wavelength for a spectral line from line profiles outside the flare region in surrounding active region and quiet Sun areas. For example, in Figure 4 the true wavelength of Fe XV should be 284.16 Å, but we measure 284.184 for the surrounding regions which we assume are at rest. Our downflow speeds are calculated using the local first order corrected wavelengths. As another example, in Figure 10 the first order local wavelength for Fe XII is the true rest wavelength, but the Fe XIV local rest wavelength is greater than the true rest wavelength by about 0.037 Å.

Physical quantities such as temperature and density are calculated from line ratios. Two lines of Fe XIV (264.78, 274.20 Å) are used to calculate electron density. The electron temperature for the multimillion degree flare lines is calculated from the ratio of the Fe XXIV 255.10 Å line to the Fe XXIII 263.76 Å line. Densities and temperatures are derived from the line ratios using the atomic data in CHIANTI (e.g., Landi et al. 2013). In the absence of a specific choice for temperature determination, the electron temperature of peak emitting efficiency of the lines has been assumed. This assumption may not be strictly valid in surge and flare plasmas if there are departures from ionization equilibrium.

The *Geostationary Operational Environment Satellite* (*GOES*) and *RHESSI* X-ray light curves for the 25 September 2011 flare are shown in Figure 1. The flare began around 15:30 UT and the three most intense hard X-ray bursts peaked at about 15:30:40 UT, 15:31:05 UT, and 15:31:40 UT. The *GOES* flux peaked at about 15:32:40 UT.

Figure 2 shows EIS raster images in the 264 Å Fe xiv line and the Fe xxiii 263 Å line. The EIS raster began at 15:26:37 UT. The EIS slit began to cross the flare region at about 15:30:33 UT, close to the time when the hard X-ray bursts began. The spectra obtained from this time to about 15:31:27 UT crossed the *RHESSI* footpoint regions. The *RHESSI* footpoint regions are the footpoints of loops into which non-thermal electrons have propagated. These electrons heat the chromosphere at the footpoints while producing thick-target hard X-ray Bremsstrahlung. As will be shown below, strong outflows due in part to chromospheric evaporation are seen in the EIS spectra recorded at 15:31:08 UT, marked by a vertical white line in the Figure 2 panels.

The speeds we measure are Doppler speeds along the line-of-sight. However, this flare was closer to the east limb than to Sun center. Thus, if the actual outflow speeds are radial, then the Doppler speeds we measure are about 0.73 times the radial speeds. This difference does not qualitatively affect any conclusions drawn from the Doppler speeds. The turbulent speeds (from line widths) are independent of the flare location.

Figure 3 shows the EIS spectra at four X-pixel locations over or close to the *RHESSI* footpoint regions. The top panels show EIS spectra as a function of Y-pixel. The bottom middle and right panels show the Fe xxiv and Fe xxiii images and *RHESSI* 25-50 keV contours. The bottom left panel shows the *RHESSI* lightcurve with vertical lines indicating the times of the four spectra shown in the top panels. The arrows in the top and middle bottom panels show the locations in Y-pixels of the main upflow regions, which are most pronounced in the X-pixel = 7 spectrum. This figure summarizes the most important EIS observations of *RHESSI* footpoint regions and evaporation signatures in multi-million degree ions. Note that the Fe xxiv line is saturated in the X-pixel = 9 image in the bright spot-like region.

The spectra at the locations of the arrows in Figure 3 show large blueshifted components. These are shown in Figure 4 for the Fe xxiv line (EIS X-pixel 7 spectrum in Figure 3). In the top panel, the histogram spectrum is composed of at least two components. There is a stationary component centered on the rest wavelength of the line (192.03 Å) and a relatively strong blueshifted component. This blueshifted component could be represented as a single Gaussian, but this is only a convenient assumption. The profile is probably composed of many unresolved regions with evaporating plasma that have slightly different outflow speeds. This would account for the large width of the Doppler shifted component. The top panel histogram spectrum corresponds to the brightest Fe XXIV region in the X-pixel 7 spectrum, indicated by the solid arrow. If the top panel histogram spectrum is represented by two Gaussians, and a temperature of 16 MK is assumed (this is typical for thermal X-ray flares), then the broad blueshifted component has a non-thermal motion of 192 km/s. The non-thermal motion for the non-shifted spectral component is 124 km/s. This is a large non-thermal motion for a stationary component. However, the EIS data occur at the time and locations of the hard X-ray burst when the flare is particularly energetic. The "average" upflow Doppler speed for the Fe XXIV emission is about 360 km s$^{-1}$, but there is significant emission at least up to 600 km s$^{-1}$.

The dashed arrow in the top panels of Figure 4 corresponds to a weaker blueshifted outflow, about 25″ south of the intense component. The solid spectrum in the top panel is for this region, and appears to be entirely blueshifted. The peak of the blueshifted emission corresponds to a Doppler speed of about 285 km s$^{-1}$, and the wing of the blueshifted emission extends to speeds greater than 500 km s$^{-1}$. Although both regions are producing outflows, it is clear that the morphology of the regions must be different. The non-thermal motions in the entirely blueshifted component represented by a single Gaussian are about 300 km s$^{-1}$ (after subtraction of the thermal and instrumental contributions). This non-thermal motion is higher than found in the X-ray spectra from *Yohkoh* and earlier missions. This is probably because EIS can isolate particular flare regions while all of the X-ray spectra were sums over the entire flare region due to their lack of spatial information.

The bottom panels in Figure 4 show spectra obtained from the same locations but for the coronal (2 MK) line of Fe XV at 284.16 Å. There is a weak blueshifted component in the histogram spectrum, and a stronger blueshifted component in the weaker spectrum, but in the weaker spectrum a component near the stationary wavelength is also present, unlike the situation for the Fe XXIV line. This component of the Fe XV line in the histogram spectrum is not really stationary. The peak emission is not at the stationary wavelength for Fe XV. The Fe XV line shows a significant downflow speed (towards the Sun) in the histogram spectrum on the order of 50 km s$^{-1}$. This downflow is seen in several coronal lines. The downflow speeds at Y-pixel = 55 for the ions, Fe X, Fe XII (the 192.39 Å line), Fe XIV, and Fe XV, are 43, 34, 35, and 49 km s$^{-1}$, respectively. These speeds are measured relative to Y-pixel = 70, which is typical for regions outside the flaring region. Note that the Y-pixel positions given above correspond to the scale in Figure 4. The Y-pixel cutout of the raster shown in the top panels of Figure 3 and in the cutouts shown in Figure 4 is from Y-pixel = 50 to Y-pixel = 120 for the entire raster. Thus the Y-pixel = 70 mentioned above is really Y-pixel 120 if the entire raster is considered, and the Y-pixel = 55 in Figure 4 is Y-pixel = 105 in the entire raster.

The Fe XXIV solid line spectrum in Figure 4 (top panel) and coronal spectra for Fe XV and three other coronal lines at the same Y-pixel location (Y-pixel = 74 on the full raster, Y-pixel = 24 on the cutout in Figure 4) are shown in Figure 5. The Y-pixel numbers in Figure 5 refer to the entire raster. They correspond to the position of the dashed arrow in Figure 4. Later we argue that the location of these spectra is at a pure footpoint region, uncontaminated by other emission along the line of sight. The coronal lines in Figure 5 do not exhibit any large downflows. The line profiles appear approximately stationary with an upflow component.

Figure 4 shows the spectra at only two Y-pixel positions. Above and below these pixels, within a few arcsec, the Fe XXIV spectrum changes substantially, indicating a complex structure over smaller scales than the spatial resolution of EIS (about 2″). This is illustrated in Appendix A. Similar complexity was found for another event discussed by Doschek et al. (2013).

So how do we interpret the spectra in Figure 4? We intepret the data as indicating that the slit is positioned over several different regions of the flare. In the histogram Fe XXIV spectrum, the line-of-sight intersects both a footpoint and a higher position in a flare loop or loops, at or near the loop-tops because this part of the spectrum is stationary. When evaporating plasma reaches near the loop-tops its upflow speed approaches zero. In contrast, the centroid of the solid line Fe XXIV spectrum is completely shifted and the spectrum seems quite symmetric and Gaussian. This is the type of spectrum expected from an evaporating footpoint. In this case, there is no ambiguity along the line-of-sight due to intersections of multiple spatial positions in a flare loop. Inspection of the spectra in Appendix A shows that along the slit there are some locations where only footpoint regions are seen, and other locations where both loop-tops and footpoints intersect the line-of-sight. The Fe XV spectra show a different behavior from Fe XXIV at the same locations as the Fe XXIV spectra. The large evaporation speeds are not seen or are weaker in the cooler coronal lines.

The AIA images at the time of the hard X-ray bursts are frequently saturated. However, there are some unsaturated images in some wavelength bands with short exposures. But what AIA does reveal in movies over the time of the hard-X-ray bursts and at later times is the eruption of a complex large surge whose origin is close to the locations of the hard X-ray bursts. The surge erupts towards the southeast as shown in AIA images in Figure 6. The images are from the 193 Å filter, which features mostly Fe XII emission but also contains the Fe XXIV line near 192 Å. The dark vertical line in the images is the position of the EIS slit at 15:31:08 UT.

The co-alignment of the EIS slit with the AIA data was produced by a cross-correlation program developed by one of us (HPW), facilitated by plotting the intensity of the Fe XII line near 192 Å as a function of Y-pixels. This is also done for the AIA image obtained at 15:30:58 UT, within seconds of the EIS image at 15:31:08 UT. The correspondence is shown in Figure 7, where the AIA data are smoothed to match better the EIS spatial resolution. The similarity of the two curves indicates that the EIS co-alignment with AIA is fairly good.

As the surge proceeds, the EIS slit continues to move eastward and snapshot spectra are captured of the surge material. Thus, the imaged plasma from the footpoint regions discussed above contains plasma evaporating in closed flux tubes as predicted by the Standard Flare Model and also material that is part of the surge. Spectra of such complex activity are rare and in this case result from the fortuitous coincidence of having a spectrometer slit over the footpoint regions produced by energetic electrons impinging on the chromosphere.

Figure 8 shows the outward moving material in Fe XIV line emission at 264.78 Å. In ionization equilibrium, Fe XIV is emitted at 2.0 MK. The left panel is an image in Fe XIV of the raster where the intensity is the integral over the entire line profile. The right panel shows the image where the intensity is the total of all wavelengths less than the rest wavelength that corresponds to Doppler outflow speeds greater than 140 km s$^{-1}$. Note from comparison with the 193 Å images in Figure 6 that most outflowing material at coronal temperatures is due to the surge. The multi-million degree flare lines do not show the surge nearly as well as the coronal lines.

The Doppler speeds in the surge material are complex. Examples are shown in Figure 9. This figure shows line profiles at various Y-pixel positions along the slit at the position of the vertical white line in the right panel of Figure 8. The component of the profile near the rest wavelength indicates a downflow speed of about 0-20 km s$^{-1}$. The blueshifted component has a range of speeds from about 0 to about 300 km s$^{-1}$ and shows a spatial variation over Y-pixels of average outflow speeds.

As another example, the surge can also be seen in Figure 10 where spectra are shown for the X-pixel = 8, Y-pixel = 87 spectrum (see left panel of Figure 8 for the location marked by the small black circle). Profiles of coronal lines in Figure 10 are completely blueshifted and the high temperature Fe XXIV line also shows a strong blueshifted component. In this figure the dot-dashed (top) and dashed (bottom) profiles are unshifted line profiles obtained from X-pixel = 8 and Y-pixel = 120. The Fe XIV line is blueshifted by about 200 km s$^{-1}$ and exhibits an extremely broad profile. Assuming that the temperature of Fe XIV is 2 MK, the non-thermal speed implied by the Fe XIV line width after correction for instrumental broadening is 140 km s$^{-1}$. The ratio of the Fe XIV 264 Å line to the Fe XIV 274 Å line is 2.95, which is in the high density limit implying an electron density greater than $10^{11}$ cm$^{-3}$. A three-Gaussian fit to the histogram spectrum in the top panel of Figure 10 reveals two Fe XXIV components, one at the rest wavelength (192.03 Å) and the other blueshifted by about 195 km s$^{-1}$, about the same result obtained for the Fe XIV line in the lower panel. In the top panel the Fe XII line is blueshifted by about 220 km s$^{-1}$. The blueshifted Fe XXIV component has a non-thermal speed of about 100 km s$^{-1}$ if the temperature is assumed to be 16 MK. The stationary Fe XXIV component has a non-thermal speed of about 90 km s$^{-1}$. The non-thermal speed for the Fe XII blueshifted line is 140 km s$^{-1}$, the same non-thermal speed deduced from the Fe XIV line in the lower panel.

There are a few other places in the spectra that appear to show outflows or downflows because extensions of emission are seen to shorter and longer wavelengths to the main emission. However, no large outflows or downflows are seen in the main profile, and we feel that the extended emission may just be continuum emission.

Electron densities can be obtained for the upflow and surge regions from ratios of the Fe XIV 264 Å and 274 Å lines. Unfortunately, densities from higher temperature ions such as Ca XIV are not available for this study. The Fe XIV density as a function of Y-pixel for the spectrum obtained at 15:31:08 UT (X-pixel = 7) is shown in Figure 11. The densities that correspond to the bright Fe XXIV regions are a few times $10^{10}$ cm$^{-3}$. Caspi et al. (2014) found similar densities for thermal flare looptops from *RHESSI* data although some M flares were closer to $10^{11}$ cm$^{-3}$.

## 2.2. *RHESSI* Results and Relationships to EIS and *SDO* Data

X-ray observations of this flare made by *RHESSI* covered the X-ray energy range from ∼5 keV to about 100 keV (Figure 1). Imaging information is also available in the same energy range with a time cadence as short as 4 s and an angular resolution as fine as a few arcseconds. These observations allow estimates to be made of the spectrum, location, and area of both the thermal and non-thermal sources as a function of time throughout the flare. This then allows the determination of the timing and magnitude of the energy deposited by the electrons into an assumed thick target and the presumed associated increase in the energy of the hottest plasma. For the non-thermal sources, the electron spectrum can be determined at energies above ∼10 keV. The measured source areas then allow the energy flux in electrons to be estimated for input to models that predict the effects of electron beam heating at flare footpoints. For the hot plasma, values can be obtained for the emission measure, temperature, and volume of plasma at temperatures in excess of ∼10 MK, allowing the thermal energy and radiated energy to be estimated for comparison with the plasma heating over all temperature ranges as measured by other instruments.

For *RHESSI* it is also interesting to examine the AIA images in the 1600 and 1700 Å filters as these indicate energy deposition in the chromosphere. They show that the flare began while *RHESSI* was still in eclipse and before there is any signal in the *GOES* soft X-ray flux. The flare begins with small bright isolated patches and spots in AIA images in the 1700 Å filter beginning at about 15:23 UT. At this time there is just a tiny patch of increased intensity. By 15:24:31 UT several spots are enhanced. By 15:26:55 UT the first spot is strongly enhanced into a small ellipsoidal patch. By 15:28:55 UT there are several bright linear regions. These bright areas continue to brighten and appear to move away from the surface in a south-east direction. This is probably the beginning of the surge and appears to coincide with the beginning of the hard X-ray impulsive phase (see Figure 1). The surge is in obvious progress by 15:31:43 UT in the 1700 Å images.

### 2.2.1. RHESSI Spectra

Spatially integrated *RHESSI* X-ray spectra were used to determine the best-fit parameters of both the thermal and non-thermal components. Following Milligan & Dennis (2009), data from the front segment of Detector #4 were used since that was found to have the finest energy resolution at this time with good sensitivity following the anneal in May 2010. Using a single detector allows all the most up-to-date instrumental corrections to be applied for the spectral analysis including the detector energy resolution and calibration, pulse pile-up, and the latest instrument response matrix that contains the probabilities that an incident photon of a given energy will produce an electrical pulse of a given amplitude from the detector segment of interest. Using the standard *RHESSI* OSPEX spectral analysis software package, the measured count-rate spectrum in each time interval was fitted with the sum of an isothermal bremsstrahlung function (*vth* using CHIANTI spectra with coronal abundances (Landi et al. 2013)) that dominates at low energies and a non-thermal thick-target photon spectrum (*thick2_vnorm*) that dominates at high energies. The best-fit parameters of the thermal function are the emission measure and temperature of the plasma emitting the X-rays, and the parameters of the non-thermal function (assumed to be a power law above some electron energy) are the flux, power-law index, and low-energy cutoff of the spectrum of the electrons that produce the observed bremsstrahlung X-rays. Since the measured count-rate spectrum is dominated at low energies by the thermal bremsstrahlung component, only an upper limit can generally be determined for the low-energy cutoff to the non-thermal electron spectrum (Holman et al. 2011). Hence, the estimate of the total energy in the accelerated electrons is a lower limit since the electron spectrum could extend to lower energies without giving any detectable evidence in the measured count-rate spectrum.

Separating the thermal and non-thermal components in the measured count-rate spectrum is not easy, and no unique solution is possible from the available information. We have found that the above analysis method gives conflicting results during the impulsive phase of this flare depending on the assumptions made about the temperature distribution of the X-ray emitting plasma. The difficulty is that the measured count-rate spectrum often shows no clear break in slope that would indicate that the thermal component would dominate at energies below the break and the non-thermal component would dominate above the break. The assumption of a single isothermal plasma adequate to fit the measured count-rate spectrum above ∼5 keV, the effective lower energy limit with the thin attenuators in place as is the case here, results in a spectrum that falls off steeply with increasing energy. The excess count rate above this single temperature spectrum at higher energies can be accounted for with the non-thermal component but it requires a low value for the low-energy cutoff to the electron spectrum and a steeper slope than if a higher temperature component were present. The resulting total energy in electrons integrated above the low-energy cutoff may then be artificially high. Analysis assuming a single temperature in each time bin resulted in the integrated electron flux peaking more than a minute after the peak in the >25 keV X-ray flux. We note from Figure 1 that the 25 keV light curve basically agrees with the *GOES* time derivative as expected from the Neupert Effect (e.g., Neupert 1968) in which the soft X-ray emitting plasma is heated by the precipitating electrons that produce the hard X-rays. The total energy in electrons integrated over the duration of the impulsive phase of the flare from 15:29 to 15:37 UT was found to be $2 \times 10^{31}$ erg, some two orders of magnitude higher than that found using the more likely assumptions discussed below.

A more realistic scenario is that the X-ray emitting plasma has a distribution of temperatures rather than just a single value (e.g., Dere & Cook 1983; McTiernan et al. 1999; Warren et al. 2013; Caspi et al. 2014). We take the next simplest assumption that there are two different temperatures, an assumption that Caspi & Lin (2010) found to fit well the data throughout the X4.8 flare on 23 July 2002. Longcope & Guidoni (2011) found two temperature components from their modeling work, with a large peak in the differential emission measure at 15 MK and a much smaller peak at 30 MK. For the *RHESSI* spectra, the low temperature component can account for the majority of the measured counts below ∼10 keV with the higher temperature component accounting for the counts at higher energies. This allows the thermal component to extend to higher energies than was possible with the single temperature assumption. The remaining higher energy counts can then be covered by photons from electrons with a higher low-energy cutoff and a flatter spectrum, both parameters that result in a lower total energy in electrons. The measured count-rate spectrum covering 16 s at the time of impulsive peak in the 25 - 52 keV count-rate light curve starting at 15:31 UT is shown in Figure 12 along with the best-fit model and all of its components folded through the instrument response matrix.

The parameters of the best-fit model shown in Figure 12 are as follows:

Thermal 1 (vth), temperature = 1.0 keV (11.6 MK), emission measure = $3.3 \times 10^{49}$ cm$^{-3}$. Thermal 2 (vth), temperature = 2.4 keV (28 MK), emission measure = $3 \times 10^{47}$ cm$^{-3}$. (The iron abundance was fixed at two times the photospheric value for both components.)

Non-thermal power-law electron thick-target component (thick2_vnorm): electron flux at 50 keV = $1.4 \times 10^{32}$ electrons s$^{-1}$ keV$^{-1}$, spectral index $\delta = 4.25$, low energy cutoff $\leq 28$ keV.

The albedo component was computed using the method described by Kontar et al. (2006) assuming isotropic emission and taking into account the location of the flare on the solar disk.

As supporting evidence for a multi-temperature model, previous observations obtained with Bragg crystal spectrometers on the Japanese *Hinotori* spacecraft reported by Tanaka (1986) sometimes showed that flares have a multithermal temperature distribution above 20 MK. Temperatures well in excess of 20 MK are the so-called "super-hot" component first reported by Lin et al. (1981). The *Hinotori* Bragg spectrometers observed the resonance lines of Fe XXV and Fe XXVI. These ions have contribution functions that peak at about 50 MK and 100 MK, respectfully. Average temperatures are obtained from dielectronic line intensities to resonance line intensities and are independent of the state of ionization equilibrium.

Analyzing the spectra with the two-temperature plus power-law assumption given above results in a peak in the non-thermal electron energy flux shown in Figure 13 at the same time as the peak in the >25 keV X-ray flux and in the *GOES* time derivative. The total energy in electrons integrated over the six time intervals analyzed and shown in Figure 13 is about $7 \times 10^{28}$ erg, almost two orders of magnitude less than the value obtained with a single temperature assumption. This is a lower limit since the low-energy cutoff found for the electron spectrum is an upper limit. The energy in electrons scales inversely with the low energy cutoff raised to the power of ($\delta$ - 2). Thus, the total energy in electrons could be increased by an order of magnitude by decreasing the low energy cutoff to 10 keV.

The parameters of the *RHESSI* thermal model can be compared with predictions from the *Hinode* X-ray Telescope (XRT). AIA and the *SDO*/EVE (Extreme Ultraviolet Variability

Experiment) data are not suitable for confirming a two-temperature thermal model. This is because none of these data give good estimates of emission measures near or greater than 20 MK (e.g., Warren 2014). However, images with the thick Be filter on XRT detect plasmas in the 20 MK range. Figure 14 shows that the XRT bright region and the *RHESSI* 6 - 12 keV image have the same location. We have calculated temperatures, emission measures, and count rates from the *GOES* data. The XRT count rate light curve looks like the *GOES* light curve except that the *GOES* light curve is lower than the XRT light curve by a factor of about 1.5. If the *GOES* light curve and emission measure are multiplied by 1.5, the XRT and *GOES* light curves are in excellent agreement. The factor of 1.5 may arise from inter-calibration and the isothermal approximation used in the analysis of the *GOES* data. At the peak of the *RHESSI* bursts (about 15:31 UT) the adjusted *GOES* (and therefore XRT) emission measure is about $2 \times 10^{49}$ cm$^{-3}$, close to the thermal *RHESSI* model prediction. However, the *GOES* temperature is about 15.5 MK, which is higher than the *RHESSI* thermal temperature. Such differences are not unexpected because a two-component flare model is only a first order approximation to the differential emission measure of a flare. The *RHESSI* high temperature component emission measure seems too small to assess with XRT data because of the dominant low temperature component.

### 2.2.2. RHESSI Images

Since *RHESSI* data are recorded for each detected photon, images can be made in any arbitrary energy and time bins providing the included number of photon counts is sufficiently high, typically greater than ∼10,000. Thus, *RHESSI* images can be made at different times during the flare in both the thermal and non-thermal energy ranges to provide the required source areas in the two spectral regimes. For the purposes of this paper, the *RHESSI* images in the non-thermal energy range allow the area of the hard X-ray footpoints to be estimated. This information, when combined with the power in electrons injected into the footpoints derived from the spectral analysis discussed above, gives the energy flux density. The total thermal energy of the hot X-ray emitting plasma can also be estimated. This requires an estimate of the volume of the plasma along with the emission measure and temperature derived from the spatially integrated spectral analysis discussed above. The volume can be determined from the area of the sources in the *RHESSI* images with some reasonable assumptions about the source extent along the line of sight.

Figure 14 shows an AIA 1700 Å image (top left panel) at the time indicated during the hard X-ray bursts with an overlay of the *RHESSI* 6-12 and 25-50 keV contours (see the lower left panel of Figure 3). The *RHESSI* 25-50 keV source has four components: a northern and southern component that are further east than what appears to be the most intense *RHESSI* source that is between the northern and southern sources, and there is another northern source west of the most intense source. Henceforth the "northern" source refers to the source east of the most intense source near X = -665 in Figure 14. A comparison of the *RHESSI* and 1700 Å AIA data indicates that the southern *RHESSI* source is a footpoint. The northern source is slightly offset from the AIA 1700 Å source, but still coincides with some 1700 Å AIA emission and is therefore probably also a footpoint region. The alignment of the northern *RHESSI* source with the brightest part of the 1700 Å image could be improved (but not completely) by a small clockwise rotation of the *RHESSI* image about the southern source. This would be within the alignment errors. In summary, there are bright (and partly saturated) 1700 Å AIA images that spatially coincide with the southern *RHESSI* source and partly coincide with the northern source (to within the few arcsec uncertainty in the alignment of the two instruments), as they should if the northern and southern *RHESSI* 25-50 keV images represent footpoint regions.

Also shown in Figure 14 are nearly contemporaneous AIA images in 131 and 193 Å filters as well as an open-thick-Be image from XRT on *Hinode* with the *RHESSI* overlays mentioned above. The northern source is bright near it in the AIA 193 and 131 images, but shifted from it by a few arcsec. To within alignment errors, they might almost coincide and represent cool coronal emission above the northern footpoint. XRT shows some weak emission close to the AIA brightenings and could represent the emission from hotter coronal plasma within the XRT passband. The southern source is only evident in the AIA 1700 image; it has no apparent counterpart in the other images. This supports the conclusion that it is a pure footpoint region. The northern source is probably partly a thermal source. These conclusions are supported by the EIS Fe XXIV spectra from X-pixel = 7, which overlays at least part of the northern and southern sources (see Figure 4). The southern Fe XXIV profile is completely Doppler shifted as expected for a pure footpoint while the northern component is a blend of shifted and unshifted components, suggesting a blend of a footpoint region with emission from plasma already evaporated into coronal loops.

The middle (and brightest) of the three 25 - 50 keV sources has bright emission near it in all the AIA channels, but in the XRT image the brightest emission is more coincident with the *RHESSI* 6-12 keV source as expected. The XRT source represents the hot thermal plasma confined to flare loops. The bright nearby AIA sources that are a little southeast of the brightest part of the XRT image and more coincident with the brightest part of the *RHESSI* 25-50 keV contour may be coronal emission that is simply near the line-of-sight of the 25-50 keV *RHESSI* source. The *RHESSI* 25-50 keV contour seems slightly offset from the 6-12 keV source. The 25-50 keV bright *RHESSI* source may be a coronal hard X-ray source rather than a footpoint source. The 25-50 keV source may be above the tops of the thermal loops seen in the XRT image, tilted by the geometry of the source so it looks offset from the XRT source. The location of the 6 - 12 keV source relative to the 25 - 50 keV source is precise since both contours are produced with the same *RHESSI* aspect information known to be accurate at the sub-arcsec level. All the *RHESSI* images in Figure 14 were made using all 9 detectors giving a nominal spatial resolution of $\sim 5''$. The 6-12 keV sources are presumably from the hot thermal plasma filling the coronal loops that also extend to the west, where other weaker sources are evident in the AIA and XRT images.

It should be kept in mind that there is often a small offset between AIA and *RHESSI* images. We have not included any offsets in Figure 14 so the apparent $\sim 3''$ differences in the E-W direction between the peak locations of the AIA and *RHESSI* thermal sources may not be real. In addition, line-of-sight effects further complicate the interpretation.

In spite of these caveats, we conclude that the southern *RHESSI* 25 - 50 keV source is non-thermal emission from footpoints, the northern source is partly from footpoints, and the bright middle source is possibly non-thermal hard X-ray emission from the corona. The areas of the three *RHESSI* northern, bright-central, and southern sources inside the 50% contours are 31, 65, and 27 arcsec$^2$ corresponding to 1.6, 3.0, and $1.4 \times 10^{17}$ cm$^2$, respectively. Thus, the total footpoint area can be assumed to be less than 6 x $10^{17}$ cm$^2$. This should be taken as an upper limit for several reasons. The central source may be at least partially a coronal hard X-ray source, the finite resolution of *RHESSI* imaging with these detectors ($\sim 5''$ FWHM) limits the ability to measure smaller source dimensions (e.g., Dennis & Pernak 2009), and albedo can make the sources appear larger (e.g., Kontar et al. 2006).

The peak thermal energy in the plasma can be estimated from the emission measure and temperature determined from the spectral fits discussed above and the volume estimated from the *RHESSI* images. The area, A, of the thermal source in the 6 to 12 keV image shown in Figure 14 at the time of peak thermal emission at $\sim$15:32 UT was determined to be $\sim 26''^2$ inside the 50% contour. If we assume that the three-dimensional source was a sphere with this equatorial a rea, then the volume would be V = 4/3 $\pi^{-0.5}$ A$^{3/2}$ = $4 \times 10^{25}$ cm$^3$. The peak thermal energy of the two thermal components if they both had this same volume can be obtained from the relation, U$_{th}$ = 3kT(EM V f)$^{1/2}$ erg, where f is the filling factor assumed = 1, and k is Boltzmann's constant. Using the temperatures and emission measures given above in Section 2.2.1 for the two thermal components in the best-fit model to the observed spectrum, we obtain the following values for the peak thermal energies. The low temperature component is U$_{th}$ = 2 x $10^{29}$ erg, and the high temperature component is U$_{th}$ = 4 x $10^{28}$ erg.

## 3. SUMMARY AND DISCUSSION

Combining the *RHESSI* and AIA results with the EIS spectra to produce an overall view of the flare is a difficult task for several reasons. The *RHESSI* and EIS data have different spatial resolutions, many of the AIA images have saturated areas, and EIS spatial resolution and temporal cadence are not sufficient to capture a complete spectroscopic image of the flare/surge.

An EIS image of an active region with a 1″ slit and 1″ step sizes is almost independent of time since features in an active region generally evolve slowly. In contrast, for the flare/surge under discussion in this paper, only parts of the active region are viewed since the step size is 5″ and the western part of the image may have evolved considerably when the eastern part of the image was recorded. Thus, bright areas do not represent a single-time view of the flare and surge, and other footpoints and loop-tops could be missed due to the 5″ step size. In this paper we have discussed a few noteworthy regions of the Fe XXIV and Fe XXIII images shown for example in Figure 3. Here are comments on other regions of the flare/surge.

In EIS X-pixel 9 in Figure 3 the brightest region is saturated in Fe XXIV but an Fe XXIII spectrum over the brightest areas reveals a profile primarily very close to the rest wavelength with only a very weak blueshift component. So this region appears mostly like a loop-top. The O VI spectra should reveal footpoint regions. However, the O VI image looks identical to the Fe X image formed at about 1 MK. Progressively higher temperature coronal lines show progressively small departures. But the strong emission in Fe XXIV that appears in X-pixel 9 is weak at the same location in the O VI and other coronal lines.

The bright Fe XXIV region in X-pixel 9 is north of the bright *RHESSI* region by about 5″. Part of the *RHESSI* emission might be a region near a loop top, but other emission may represent footpoints given the high energy band used for the *RHESSI* images. The most southern bright *RHESSI* source in Figure 3 corresponds to the EIS footpoint region we have discussed in spectrum X-pixel 7. The other bright *RHESSI* source is a mixture of footpoint and loop top regions and has also been discussed above. We conclude that the spatial resolution of both *RHESSI* and EIS results in averaging over multiple loop top and footpoint regions in a flare arcade.

We have found evaporation or surge generating footpoints at multiple locations during the hard X-ray burst. A conclusion is that the energetic particles were not all confined to a single loop, but instead followed field lines that terminated at different locations in the chromosphere. This result is consistent with multiple flaring loops in a loop arcade, a typically observed scenario. The result is also consistent with predictions of magnetic island reconnection models (e.g., Drake et al. 2006, 2013). In these models, the electrons are accelerated by different islands and attach themselves to nearby field lines that do not all terminate at the same location. Furthermore, the energy deposited at each footpoint can be different, e.g., Figure 9 shows that the upflow speeds in the surge differ at different positions along the slit. The spectra in the Appendix also show different profiles at neighboring pixel positions along the slit.

A recent investigation of fine-scale spatial structures in coronal rain carried out with the Swedish 1-m Solar Telescope/Crisp Imaging Spectro-Polarimeter and AIA images also support the conjecture that what looks like a single magnetic flux tube at low resolution is really composed of a multitude of magnetic strands that can have 100 km cross-section widths or less (e.g., Scullion et al. 2014). Their work includes post-flare loops. From the Scullion et al. (2014) work and modeling of the reconnection region, it is expected that flare footpoint regions will be complex and not fully resolved by instruments like EIS with about $2''$ spatial resolution.

In one upflow region plasma near 13-16 MK had a line profile that peaked near Doppler speed 250 km s$^{-1}$ and extended from about zero speed to Doppler speeds greater than 600 km s$^{-1}$. Such a large line width is consistent with multiple unresolved evaporation regions that have multiple outflow speeds. The evaporation region we have discussed extensively (Y-pixel = 25, X-pixel = 7 in Figure 3) even has outflows in coronal ions such as Fe x. The predicted downflows (e.g., Milligan & Dennis 2009) at coronal temperatures are also present (about Doppler 30 km s$^{-1}$).

For coronal ions, the Doppler outflow speeds measured along the white vertical line in Figure 8 vary from Doppler downflows of about 20 km s$^{-1}$ to Doppler outflow speeds of about 300 km s$^{-1}$. Figure 10 also shows Doppler shifted surge Fe xiv ions that have non-thermal motions of about 140 km s$^{-1}$. In addition Fe xxiv is seen moving outward with Doppler speeds in the 200 km s$^{-1}$ range.

The electron density behavior shown in Figure 11 determined from the Fe xiv line ratio is puzzling. It does not show spikes of increased density along the Y-axis (X-pixel = 7, Figure 11) where footpoint regions are encountered, i.e., Y-pixels = 75 and 105 (Y-pixels = 25 and 55 in Figure 3. (The narrow spike near Y-pixel = 22 in Figure 11 is not real.) Spikes might be expected because the top part of the chromosphere is evaporated away, revealing coronal emission at lower altitudes and presumably higher densities. However, the Fe xiv density (about 2 MK) simply increases over the whole region of the flare and surge, reaching peak values of about 2x10$^{10}$ cm$^{-3}$. This might be due to line-of-sight effects or closely spaced multiple footpoint regions.

As mentioned, the complexity of this flare prevents detailed modeling without more information on footpoint locations and the flare morphology. However, we have modeled a single flare flux tube with heating input from our RHESSI data just to see if the velocities we measure from the EIS data are at least comparable to what is observed.

The numerical simulations were performed with the HYDRAD code (Bradshaw & Cargill 2013), which solves the one-dimensional hydrodynamic equations of conservation of mass, momentum, and energy for an isolated magnetic flux tube and a multi-fluid plasma (e.g., electrons, ions, neutrals). The equations and assumptions are explained in the appendix of Bradshaw & Cargill (2013). As detailed in Reep et al. (2013), HYDRAD now uses a more realistic chromosphere, based on the VAL C model (Vernazza et al. 1981), with a non-uniform ionization structure. The effects of neutral atoms on the energy balance between the two fluids is now treated in full, while the radiative energy balance is based on the recipe derived by Carlsson & Leenaarts (2012). The loops are assumed to be semi-circular, along the field-aligned direction, with a constant cross-sectional area. The initial temperature and density profiles are found by integrating the hydrostatic equations from the chromosphere to the apex of the coronal loop. Initially, it is assumed that the electron and ion populations are in thermal equilibrium, before significant heating events occur.

Because this single loop model only considers heat input from non-thermal particles, it cannot reproduce flare decay times seen in *GOES* data since no heat source is available after cessation of the hard X-ray bursts. Additional heating, often required in flares, would extend the flare thermal emission as seen by *GOES* in comparison with the modeling results. One further complication is that multi-threaded models require further assumptions such as the duration and magnitude of heating on each thread, and even the number of threads composing the flare arcade.

Another difficulty encountered with even a single loop model is the uncertainty in the electron energy flux (in erg cm$^{-2}$ s$^{-1}$) going into the chromosphere that forms the basic input to the model. This flux is the total energy of the accelerated electrons divided by the area of the footpoints. The total energy in non-thermal electrons derived from the *RHESSI* X-ray spectra is essentially a lower limit because of the strong thermal X-rays that dominate the spectrum at lower energies. The footpoint area is uncertain due to *RHESSI's* several arcsecond spatial resolution and to the fact that EIS did not observe all the footpoints because the raster scan missed parts of the flare. A final problem is that the brightest *RHESSI* source might be a coronal hard X-ray source not produced by models such as HYDRAD. We must conclude that for these reasons the total non-thermal energy and footpoint areas are highly uncertain.

The HYDRAD model was run for several initial electron densities and included 11 EIS spectral lines. The initial loop length was estimated to be 28 Mm from AIA images.

Initial model runs using the *RHESSI* parameters described in Section 2 failed to produce a reasonable agreement between the predicted *GOES* flux and the observations. It was necessary to increase the total non-thermal electron energy found in Section 2 by an order of magnitude to get reasonable agreement in the impulsive phase. This can be accomplished in the *RHESSI* analysis by lowering the low energy cutoff from 28 keV to 10 keV. This is still consistent with the *RHESSI* data because the exact low energy cutoff energy cannot be determined. Thus, a low energy cutoff of 10 keV and a total non-thermal flare energy of 7 x $10^{29}$ ergs were assumed rather than the total energy given in Section 2.

The model reproduces only the very initial rise of the *GOES* light curves. But it does reproduce the velocities of the high temperature lines from ions such as Fe XXIV, and the general absolute magnitudes of the coronal velocities were of the same order magnitude as the observed coronal velocities. However, the detailed behavior of the coronal line blueshifts and redshifts showed contradictions to the model that are not all easy to explain. Some of the contradictions can be explained by the fact that a single loop model is not appropriate for this event. Also the observed line widths of the high temperature lines are much wider than the simulated widths. We conclude that for events such as the one under discussion, multi-loop modeling is essential.

The data for the flare and surge presented in this paper show how complex flaring regions can be. EIS has revealed a plethora of activity at multiple locations at the time of the hard X-ray bursts. Current imaging and imaging spectroscopic instrumentation lack the spatial and some lack time resolution to study individual footpoint regions in flares except for the very simplest events. Dynamic range is also a considerable problem. Numerical simulations of flares are generally qualitatively consistent with the observations, but a quantitative link between a hard X-ray producing electron beam deposition in a single loop and an observed footpoint in that loop in the EUV is not possible to find for most events. The full investigation of the Standard Flare Model for a large flare requires a dedicated solar flare mission with improved instrumentation to offset these limitations. Currently we have tantalizing observations and modelling that qualitatively support the Standard Flare Model, but not to the rigor desirable. Observations such as presented in this paper should guide the design and development of future dedicated flare instrumentation.

Here are some instrumental improvements that would help in detailed testing of the Standard Flare Model:

1 - Broad-band filter images (i.e., like AIA) that do not saturate during the impulsive phase and have less than 1 s cadence.

2 - Much higher time resolution (less than 1 s cadence) EUV spectroscopic observations with spectral resolution similar to that of EIS but with more comprehensive spatial coverage than its picket-fence observations. It should be noted that higher spatial resolution observations of chromospheric evaporation in flares is currently available for some spectral lines from the *Interface Region Imaging Spectrograph* (*IRIS*), e.g., the Fe XXI line near 1354 Å (e.g., Tian et al. 2014; Young et al. 2015; Polito et al. 2015). They observe upflows in Fe XXI at footpoints and also some downflows as loops collapse from the reconnection region. However, as applies to EIS, more time resolution and spatial coverage is desirable, as well as more spectral coverage.

3 - Higher spatial resolution hard X-ray observations around the 25-50 keV energy range with improved sensitivity and dynamic range.

4 - Improved ability to co-align images from different instruments and spacecraft at the arcsec level.

5 - The highest temperature ion observable with EUV instrumentation is Fe XXIV. Therefore, high spatial and spectral resolution soft X-ray imaging instrumentation is needed to observe the superhot flare plasma near the reconnection region, in particular the Fe XXV and Fe XXVI line groups near 1.8 Å. The Fe XXVI Lyman-$\alpha$ doublet has its peak emission at about 100 MK.

6 - Spatial resolution at the sub-arcsec level for all the instruments.

*Hinode* is a Japanese mission developed and launched by ISAS/JAXA, collaborating with NAOJ as a domestic partner, and NASA (USA) and STFC (UK) as international partners. Scientific operation of the *Hinode* mission is conducted by the *Hinode* science team organized at ISAS/JAXA. This team mainly consists of scientists from institutes in the partner countries. Support for the post-launch operation is provided by JAXA and NAOJ, STFC, NASA, ESA (European Space Agency), and NSC (Norwegian Space Center). We are grateful to the *Hinode* team for all their efforts in the design, build, and operation of the mission.

CHIANTI is a collaborative project involving the US Naval Research Laboratory, the Universities of Florence (Italy) and Cambridge (UK), and George Mason University (USA).

GAD and HPW acknowledge support from the NASA *Hinode* program and from ONR/NRL 6.1 basic research funds. JWR was supported by NASA Headquarters under the NASA Earth and Space Science Fellowship Program (Grant NNX11AQ54H). AC was supported by NASA grant NNX12AH48G. We would like to especially acknowledge and thank the referee for a careful and detailed reading of the manuscript, whose comments and suggestions have considerably improved the original version of this paper.

## 4.   APPENDIX A

As mentioned in the text, the spectra of Fe XXIV for the outflowing regions change substantially as a function of Y-pixel. This is shown in Figure 15, Figure 16, Figure 17, and Figure 18 for Y-pixels involved in the lower upflow marked by an arrow at Y-pixel = 24 in the top panels of Figure 3 and also marked by the dashed arrow in Figure 4. The Y-pixels in the Appendix figures equal the Y-pixels in the top panels of Figure 3 and the cutouts in Figure 4 + 50. Thus Y-pixel = 74 on the full raster and Y-pixel = 24 on the cutouts in Figure 3 and Figure 4. The vertical lines in the Appendix figures mark the rest wavelength (192.03 Å) of the Fe XXIV line.

A comparison of the spectra reveals that at some Y locations there is a stationary and blueshifted component along the line-of-sight. Thus, in these cases there is a mixture of footpoint and loop-top emission. In other locations a completely shifted Fe XXIV profile is observed. So at these locations the emission is dominated by footpoints. The Fe XXIV lines are quite broad, probably due to the superposition of multiple upflows with slightly different velocities at a spatial resolution less than about $1''$. The spikes seen in some of the profiles (e.g., top panels of Figure 16) are too narrow to be real. They are simply due to bad data at these points in the profiles.

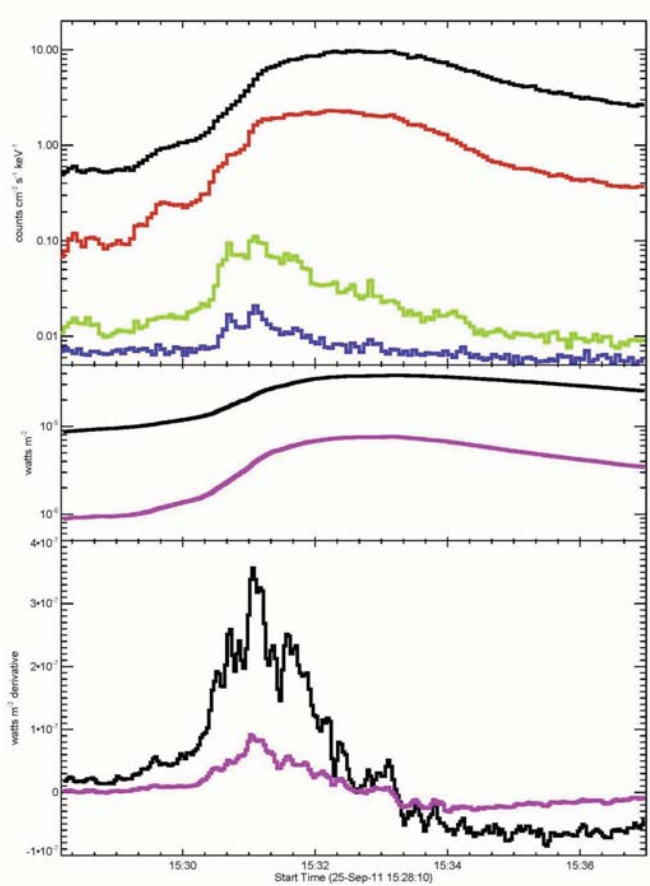

Fig. 1.— *RHESSI* and *GOES* light curves covering the impulsive phase of the flare on 25 September 2011. Top plot: *RHESSI* count fluxes vs. time for four energy bands: 6–12 keV (black), 12–25 keV (red), 25–50 keV (green), and 50–100 keV (blue). The counts were summed over detectors 1, 3, 4, 5, 8, and 9. The thin attenuators were in the detector fields of view during the plotted time interval. Middle plot: *GOES* light curves: 1–8 Å (black) and 0.5–4 Å (lavender). Bottom plot: Time derivative (3-point Lagrangian interpolation) of the *GOES* light curves with the same color code.

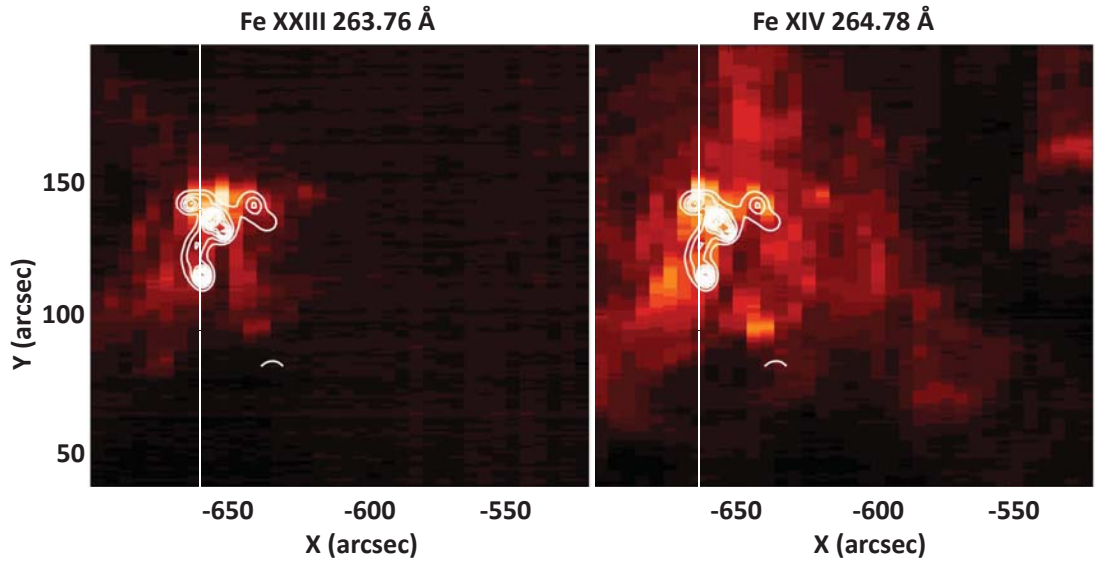

Fig. 2.— Right: The complete EIS raster in the spectral line of Fe XIV. Left: The Fe XXIII EIS raster truncated in the west limb region. The vertical white lines are the position of the EIS slit at 15:31:08 UT. The white contours (at 10-90% of the peak flux) show the *RHESSI* image in the 25-50 keV energy band.

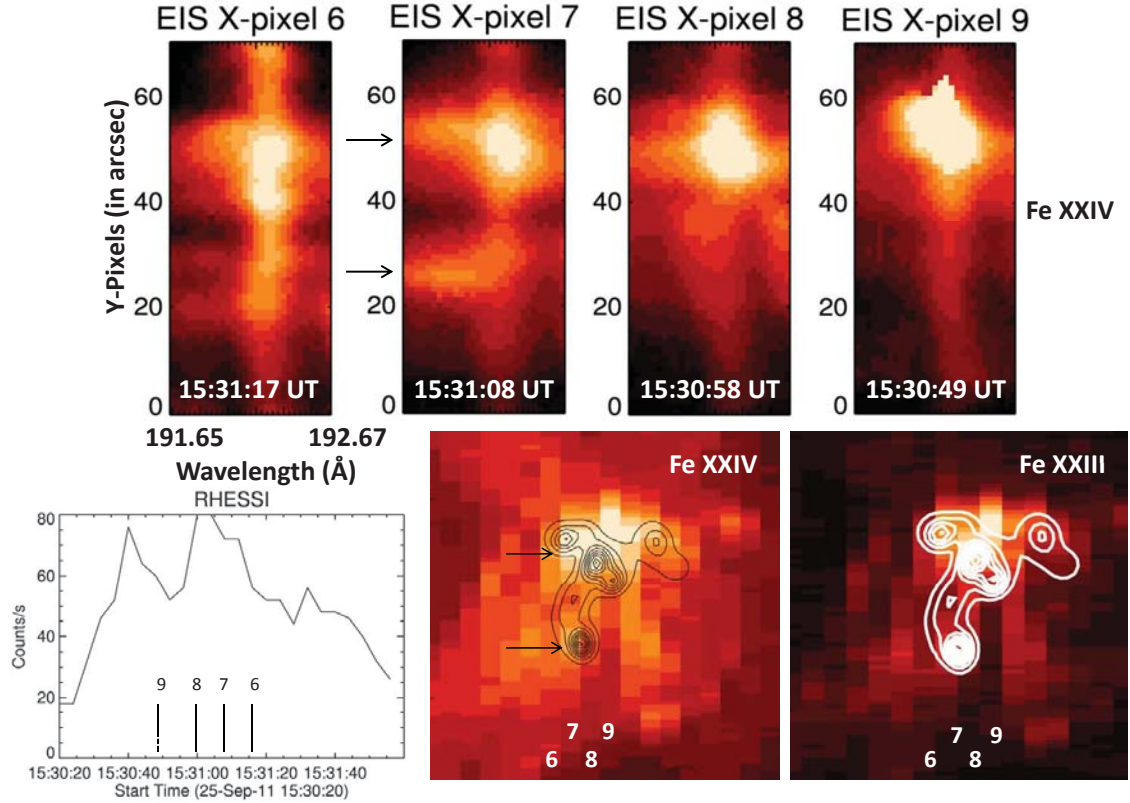

Fig. 3.— Top panels: EIS Fe XXIV spectra at the specified times as a function of Y-pixels for four X-pixel locations. Bottom middle and right panels: images in lines of Fe XXIV (192.03 Å) and Fe XXIII (263.76 Å). The locations of the four spectra are indicated by X-pixel numbers, and the arrows correspond to upflow regions seen in the X-pixel 7 spectrum in the top panels. Bottom left: the lightcurve of the 25-50 keV *RHESSI* data; the four vertical lines correspond to the times of the four spectra in the top panels. The spectra in the top panels are Y-pixel segments of the entire raster between raster Y-pixel = 50 and raster Y-pixel = 120.

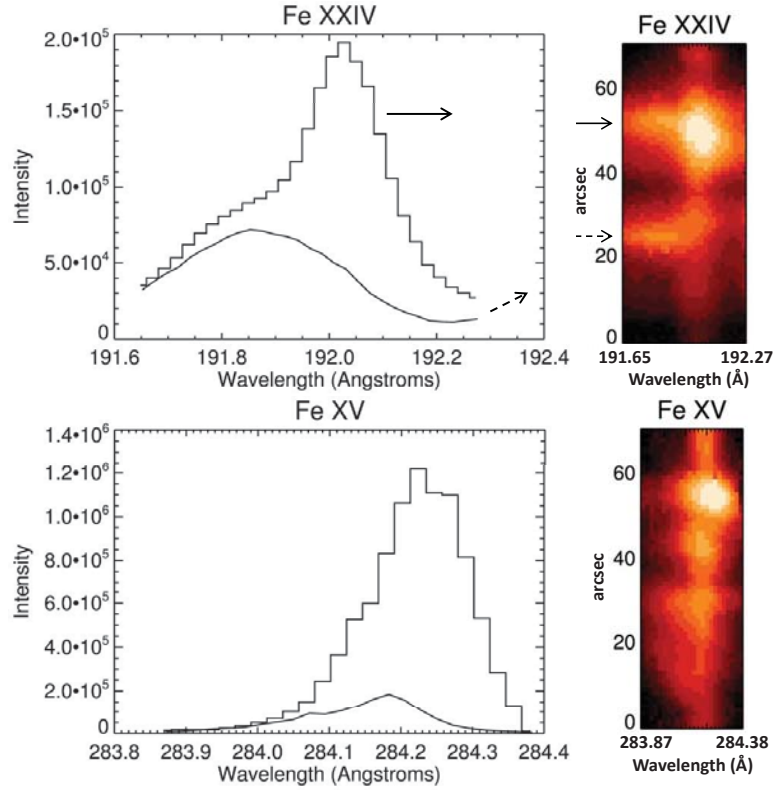

Fig. 4.— Top left: Fe XXIV spectra at the locations in the right panel indicated by the arrows. Bottom spectra of Fe XV correspond to locations close to those in the top panels. The spectra are for X-pixel = 7 in Figure 3. The Fe XXIV rest wavelength is 192.03 Å. The Fe XV rest wavelength is about 284.18 Å. The spectral images are Y-pixel segments of the entire raster between raster Y-pixel = 50 and raster Y-pixel = 120.

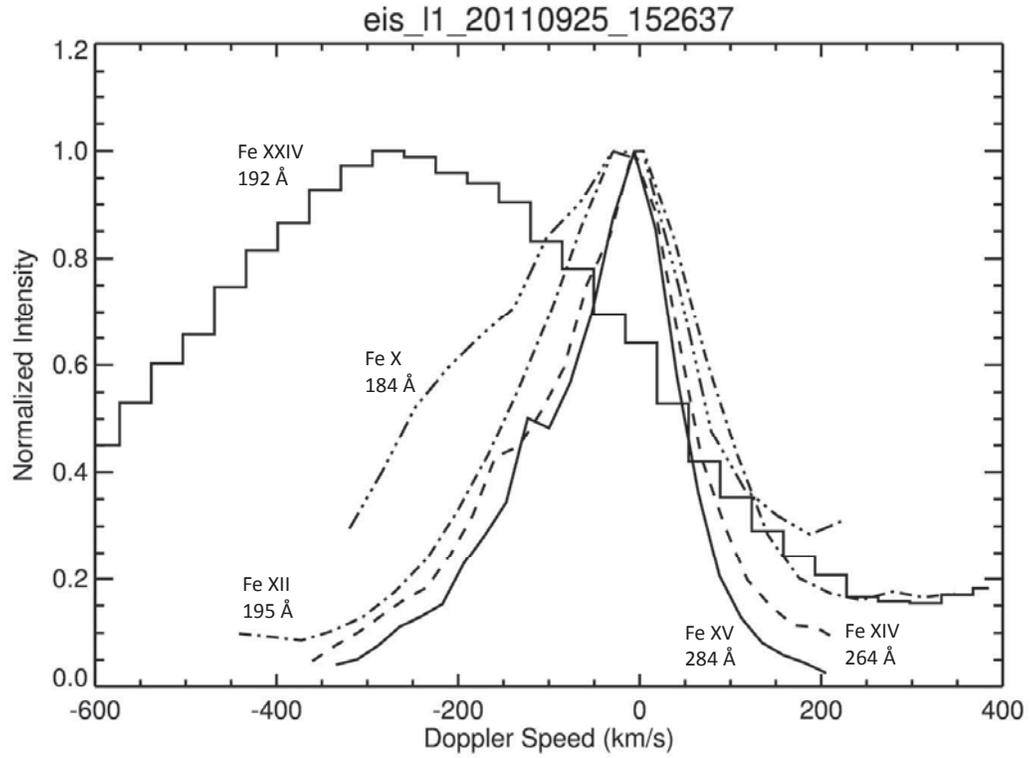

Fig. 5.— Normalized intensity observed line profiles with EIS (X-pixel = 7, Y-pixel = 74, location of dashed arrow in Figure 4) showing Doppler shifts and line widths.

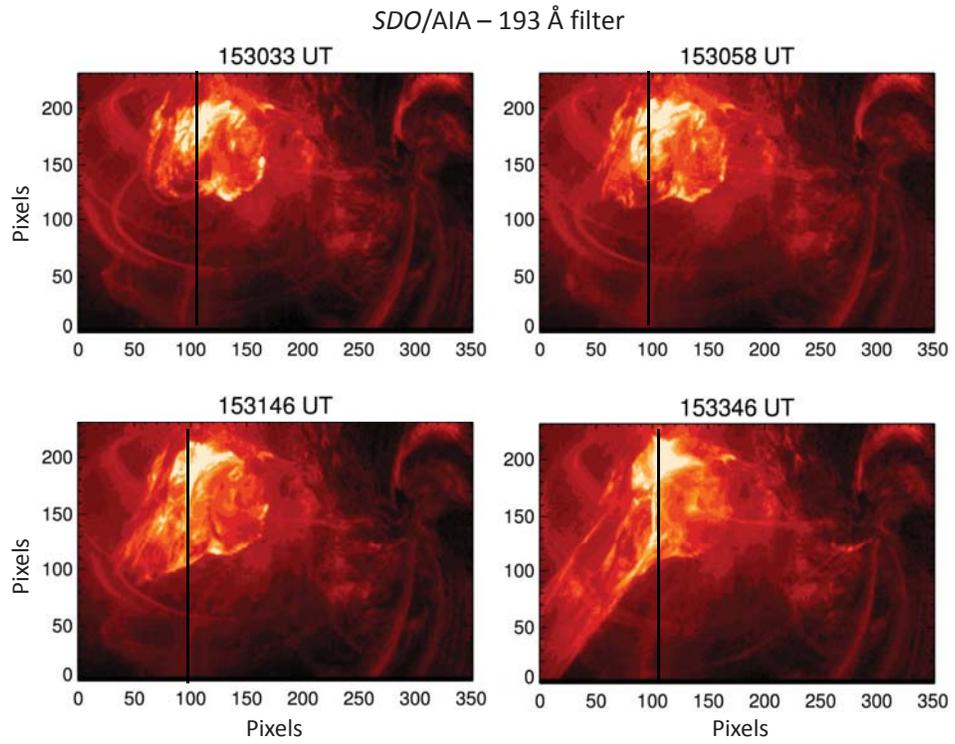

Fig. 6.— AIA images of the flare region at the times shown. The dominant emission is from Fe XII but there is also some emission from Fe XXIV. The dark vertical line is the position of the EIS slit at 15:31:08 UT.

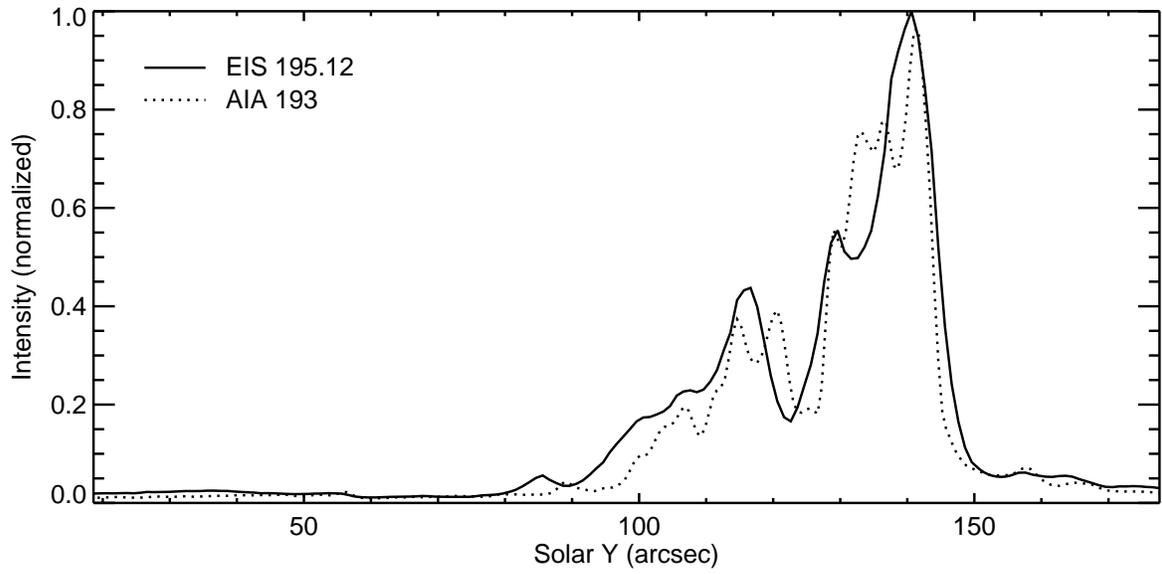

Fig. 7.— Co-alignment of EIS and AIA at 15:31:08 UT. The X-axis is Y-pixels. The general shapes of the curve are similar, indicating a good relative alignment of the EIS raster with the AIA images. The Y-pixels in this figure equal the Y-pixels in Figure 3 and Figure 4 plus 50.

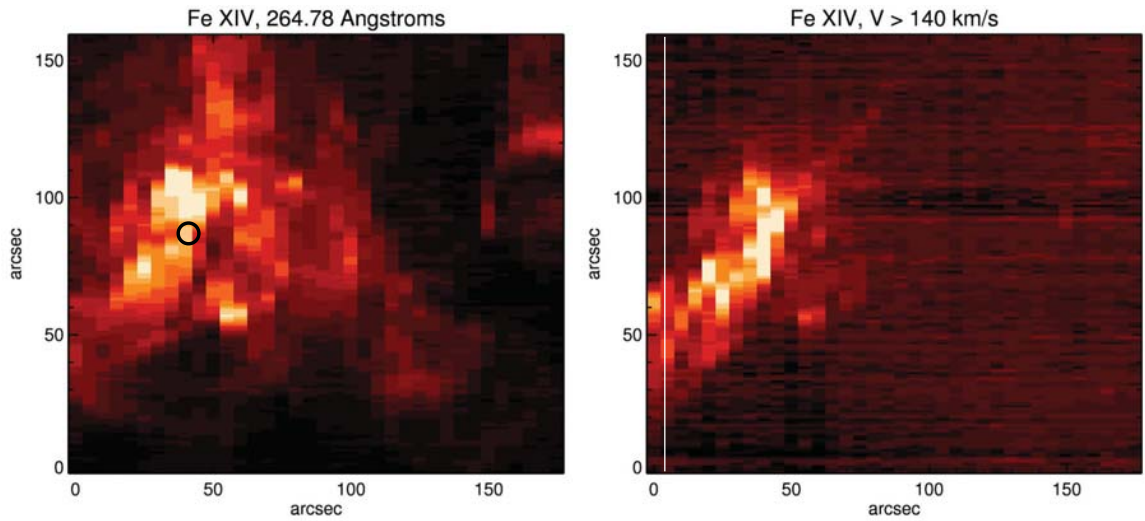

Fig. 8.— Left: The Fe XIV raster image where the intensity is the integral over the entire line profile. Right: The Fe XIV raster image where the intensity is only integrated over wavelengths corresponding to approaching Doppler speeds greater than 140 km s$^{-1}$. The white vertical line shows the slit position for the spectra in Figure 9. The small black circle refers to a surge spectrum shown in Figure 10.

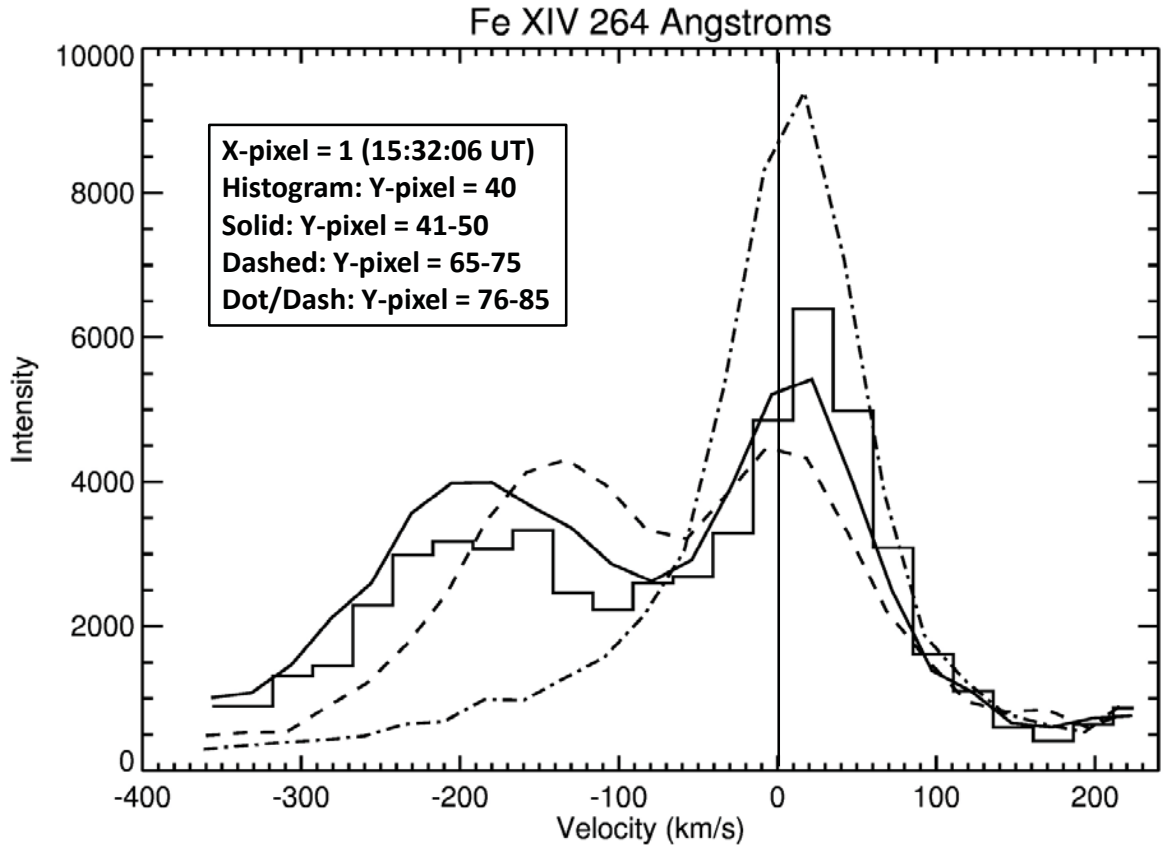

Fig. 9.— Doppler speeds along the slit for selected pixels at the position indicated by the vertical white line in the right panel of Figure 8. The spectra are averages over the Y-pixel range indicated.

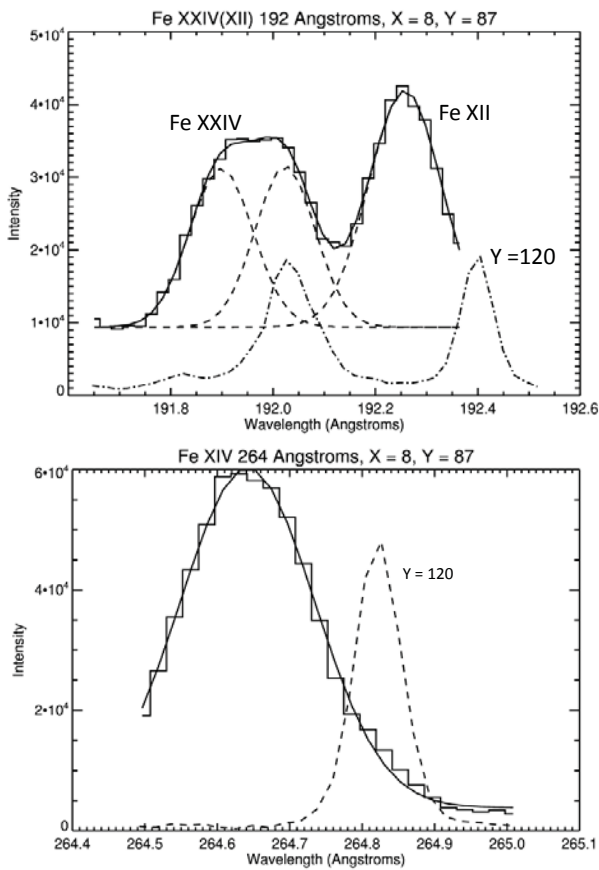

Fig. 10.— Footpoint and surge spectra at 15:30:58 UT. The coronal lines such as the Fe XIV 264 Å line are Doppler blueshifted by about 200 km s$^{-1}$ and have non-thermal motions of about 140 km s$^{-1}$. A three-Gaussian fit was made to the data in the top panel. The individual components are shown as dashed curves and the solid curve is the total fit to the data (histogram). Non-Doppler shifted spectra (dot-dashed) are shown for comparison, obtained at X-pixel = 8 and Y-pixel = 120. The Y-pixels in this figure equal the Y-pixels in Figure 3 plus 50. The rest wavelengths of the lines are: Fe XXIV, 192.03 Å; Fe XII, 192.39 Å; Fe XIV, 264.78 Å.

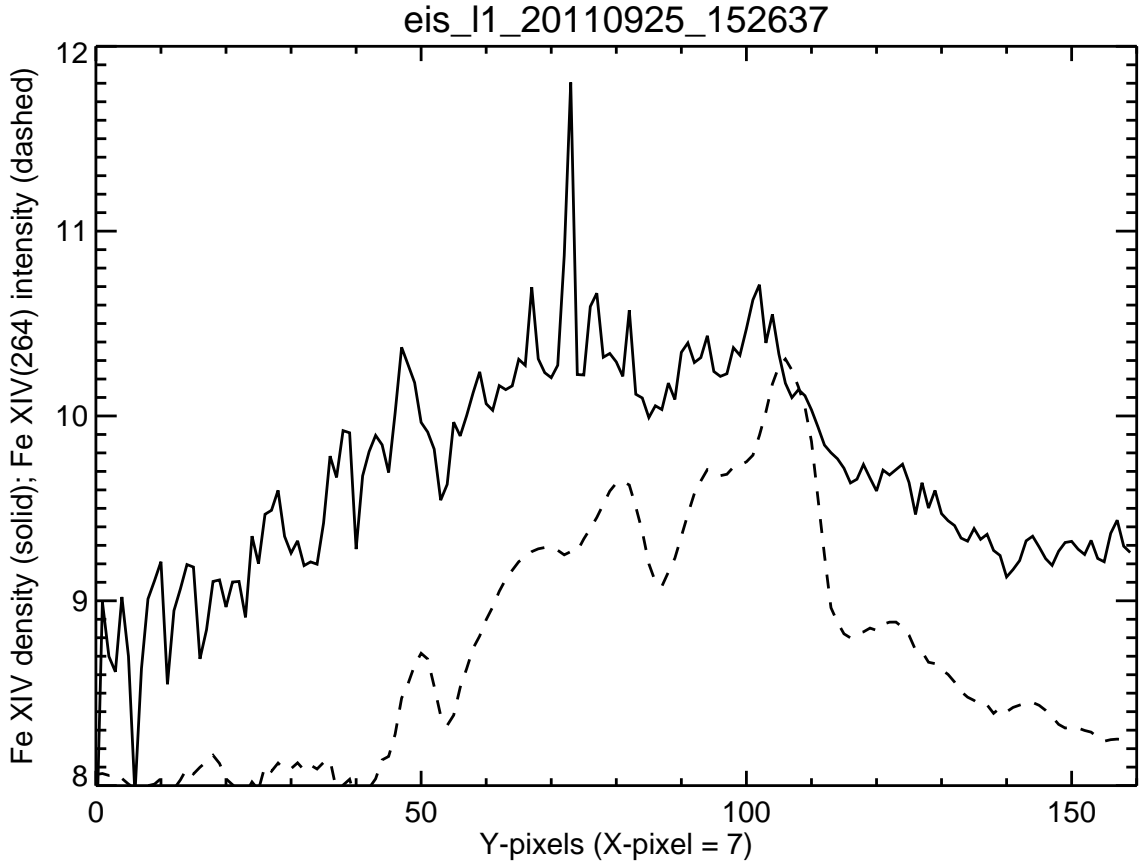

Fig. 11.— Solid line: the $\log_{10}$ electron density (cm$^{-3}$) from the Fe XIV 264/274 Å line ratio as a function of Y-pixels for the EIS spectrum obtained at 15:31:08 UT. Dashed line: the $\log_{10}$ intensity of the 264 Å line arbitrarily scaled to fit onto the density plot. The Y-pixels in this figure equal the Y-pixels in Figure 3 and Figure 4 plus 50.

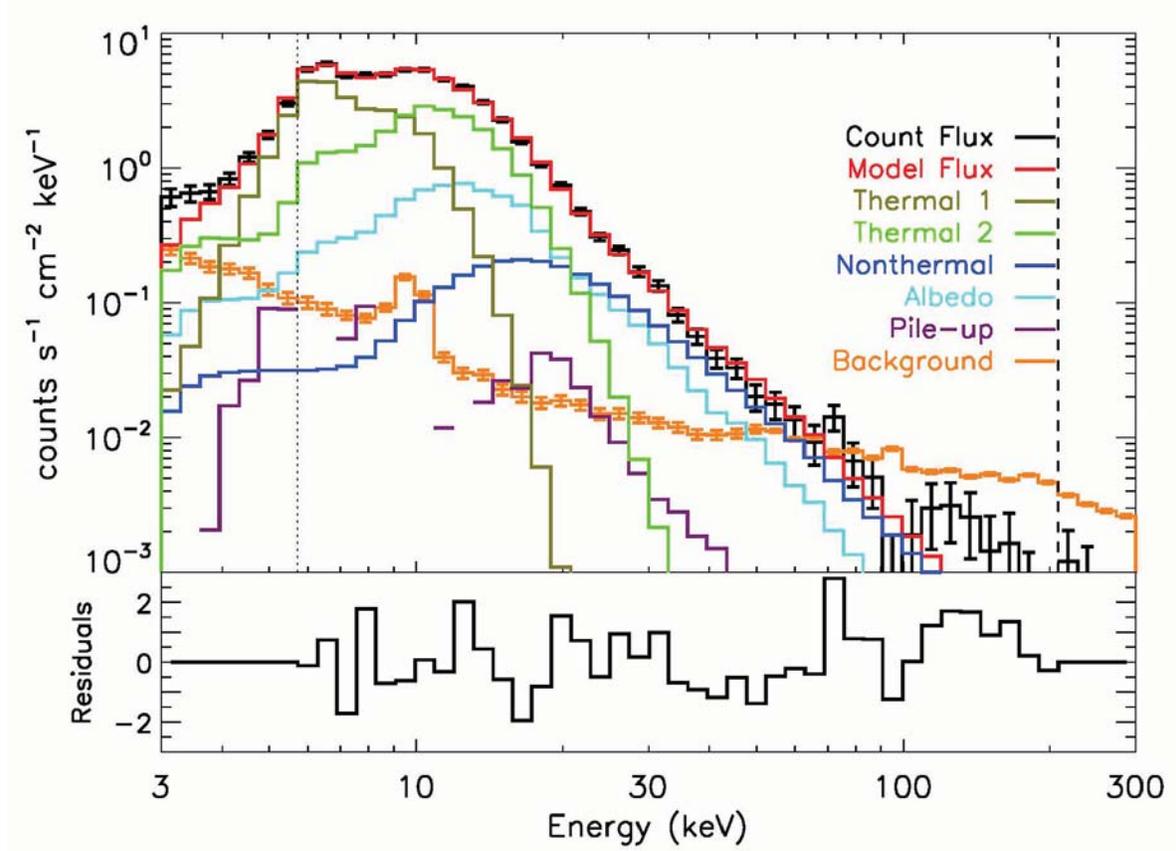

Fig. 12.— Spatially integrated, background-subtracted, count-flux spectrum measured with *RHESSI* Detector Number 4 for the 16s interval starting at 15:31 UT shown in black with $\pm 1\sigma$ statistical error bars. The model spectrum fitted between 5.8 and 210 keV and folded through the instrument response matrix is shown in red. The different components of the model are as follows: two thermal bremsstrahlung functions (*vth* shown in olive and green with the Fe abundance fixed at two times the photospheric value), a non-thermal thick-target spectrum from a power-law electron spectrum (*thick2_vnorm* shown in blue), albedo assuming isotropic emission (shown in cyan), and pulse pileup (in purple). The background spectrum subtracted from the total count-flux spectrum is shown in orange with $\pm 1\sigma$ statistical error bars. Bottom: differences between the measured and the model-predicted count fluxes normalized in each energy bin to the statistical $1\sigma$ uncertainty.

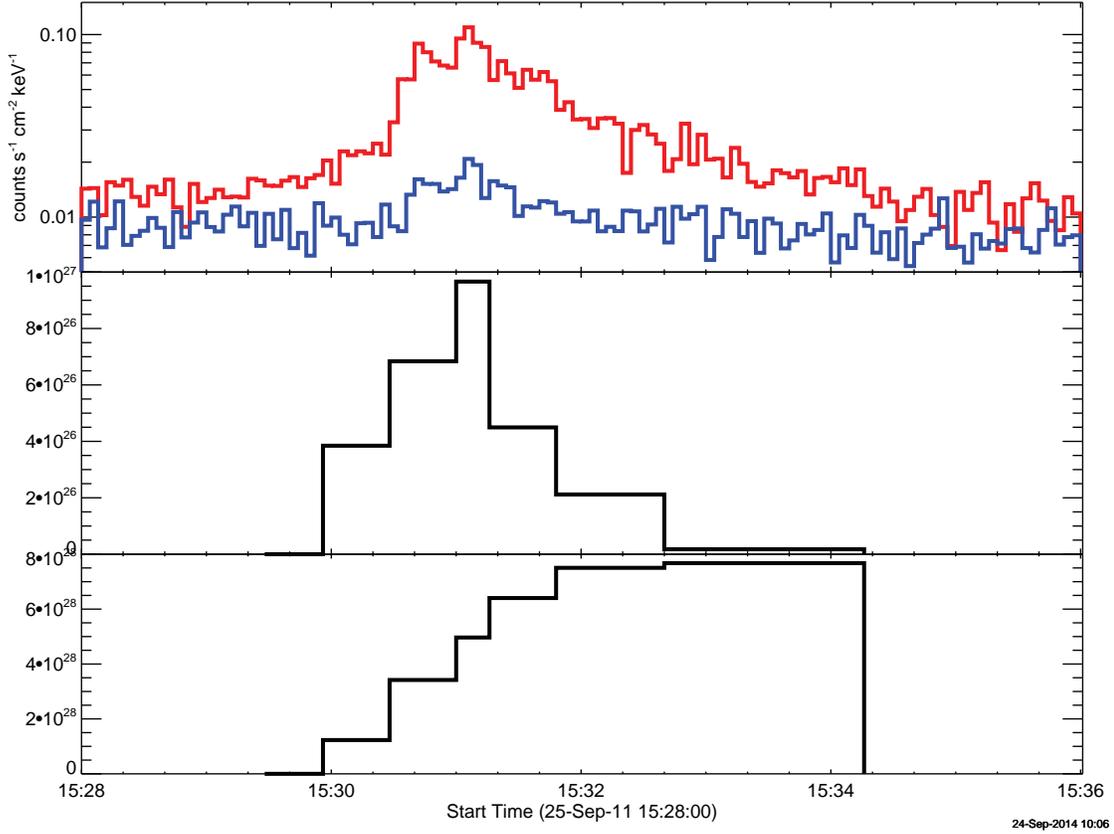

Fig. 13.— Top: Time profile of the *RHESSI* 25 - 50 (red) and 50 - 100 keV (blue) flux in counts s$^{-1}$ cm$^{-2}$ keV$^{-1}$. Middle: Estimated energy flux in erg s$^{-1}$ of electrons at energies above a variable low-energy cutoff. Bottom: Running time integral of the electron power shown in the middle plot. This gives the cumulative energy in ergs injected into the thick target as a function of time. The last point shows that a total of about $7 \times 10^{28}$ erg of energy in accelerated electrons was injected into the thick target during the flare assuming a low energy cutoff of 28 keV. The time bins for the energy estimates were chosen to give ample statistics so that the thermal and non-thermal spectral components could be more clearly separated.

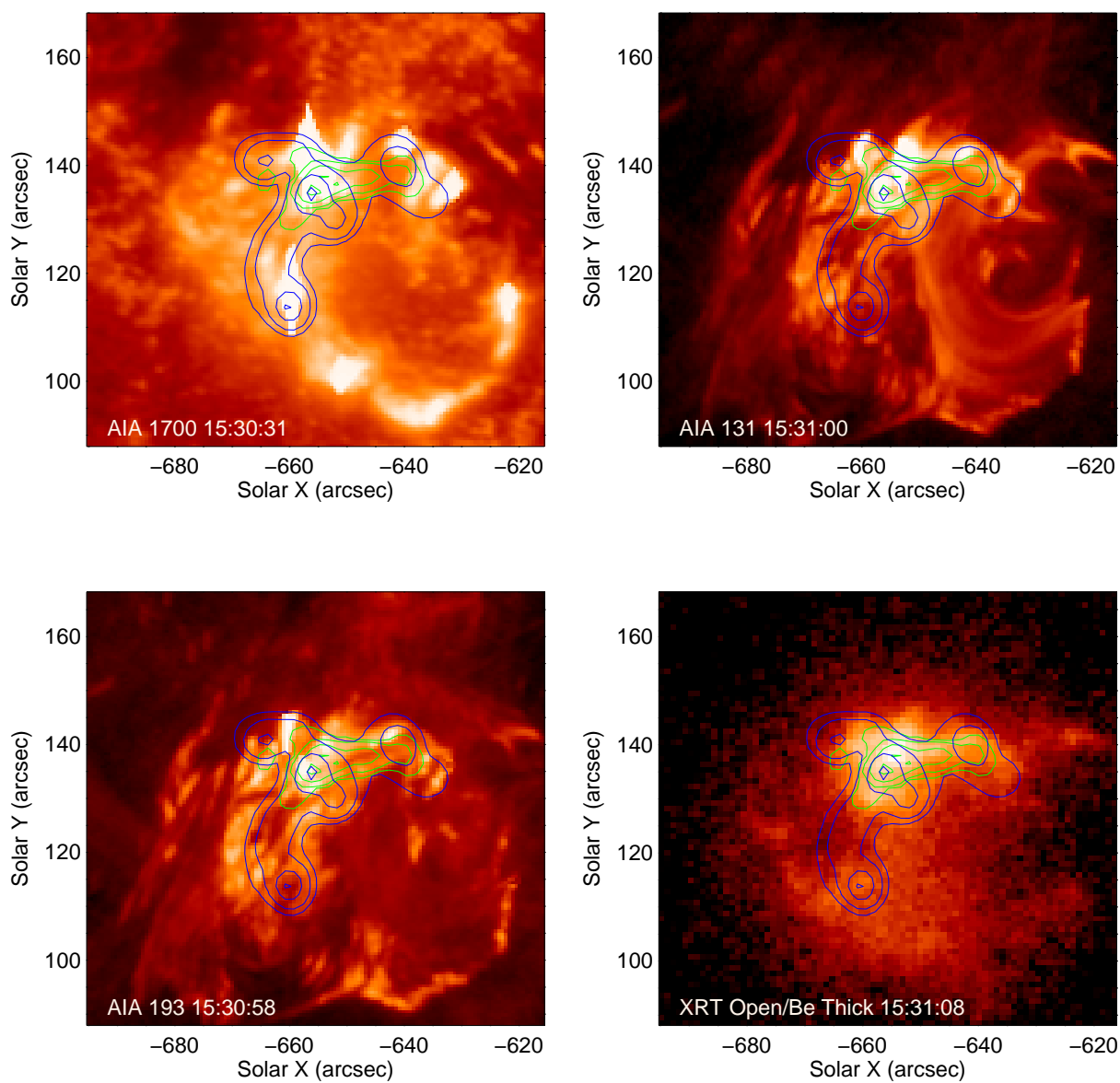

Fig. 14.— *SDO*/AIA 1700, 193, and 131 Å images and an XRT image taken close to 15:31 UT overlaid with *RHESSI* 6 - 12 (green) and 25 - 50 keV (magenta) contours obtained at about the same time. The *RHESSI* images were reconstructed with the Pixon method using data from all 9 detectors between 15:30:28 and 15:31:56 UT to include the impulsive peaks seen above 25 keV. The contours are at 10, 20, 50, and 90% of the peak value for each energy range with an extra 5% contour for the 25 - 50 keV image.

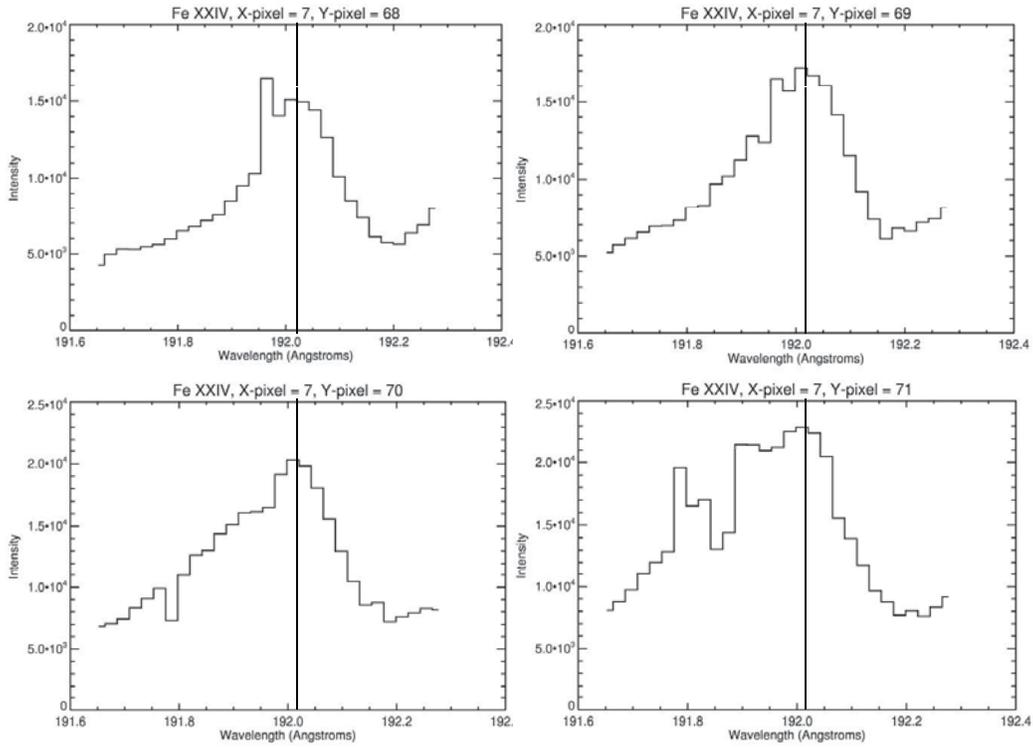

Fig. 15.— Appendix Figure 1: The Fe XXIV line as a function of entire raster Y-pixel for the upflow region marked by the dashed arrow in Figure 4. The Y-pixels in all the Appendix figures equal the Y-pixels in the cutouts of Figure 4 + 50. This is also true for the top panels of Figure 3. The vertical lines in the Appendix figures mark the rest wavelength (192.03 Å) of the Fe XXIV line.

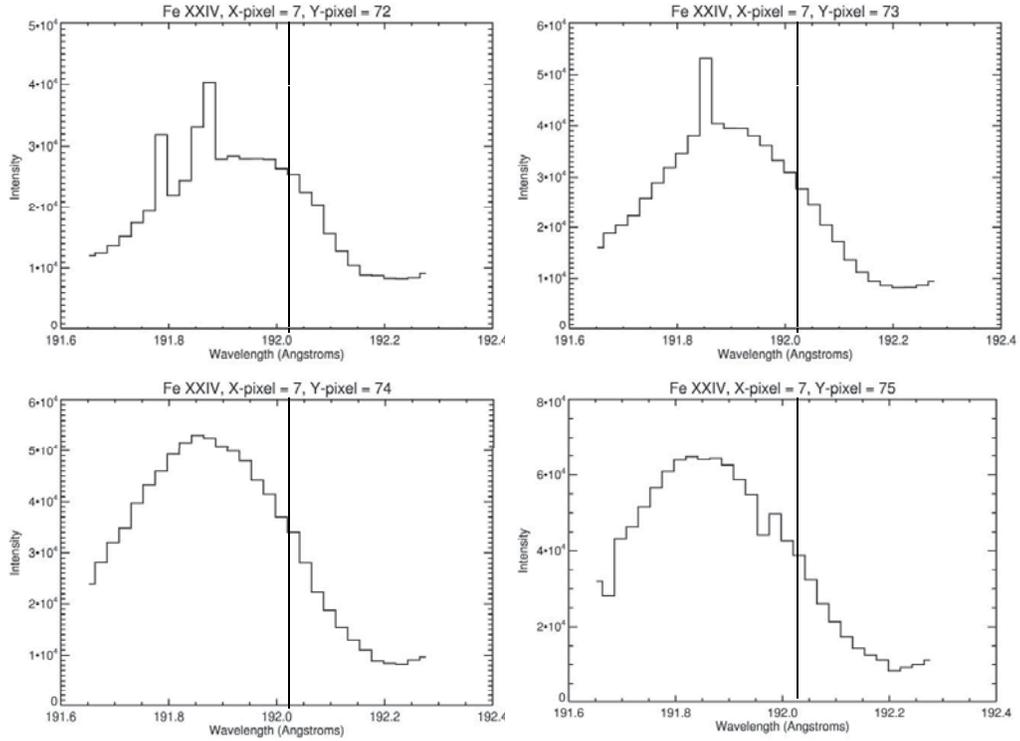

Fig. 16.— Appendix Figure 2: The Fe XXIV line as a function of entire raster Y-pixel for the upflow region marked by the dashed arrow in Figure 4. The Y-pixels in all the Appendix figures equal the Y-pixels in the cutouts of Figure 4 + 50. This is also true for the top panels of Figure 3. The vertical lines in the Appendix figures mark the rest wavelength of the Fe XXIV line. The spikes in the top two panels are not real.

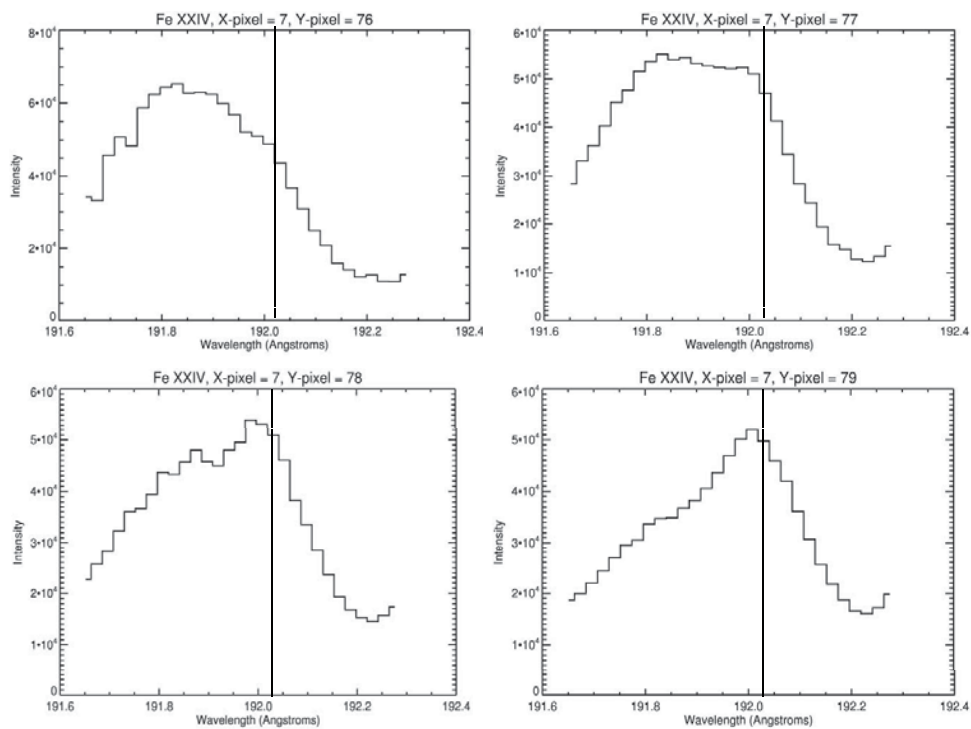

Fig. 17.— Appendix Figure 3: The Fe XXIV line as a function of entire raster Y-pixel for the upflow region marked by the dashed arrow in Figure 4. The Y-pixels in all the Appendix figures equal the Y-pixels in the cutouts of Figure 4 + 50. This is also true for the top panels of Figure 3. The vertical lines in the Appendix figures mark the rest wavelength of the Fe XXIV line.

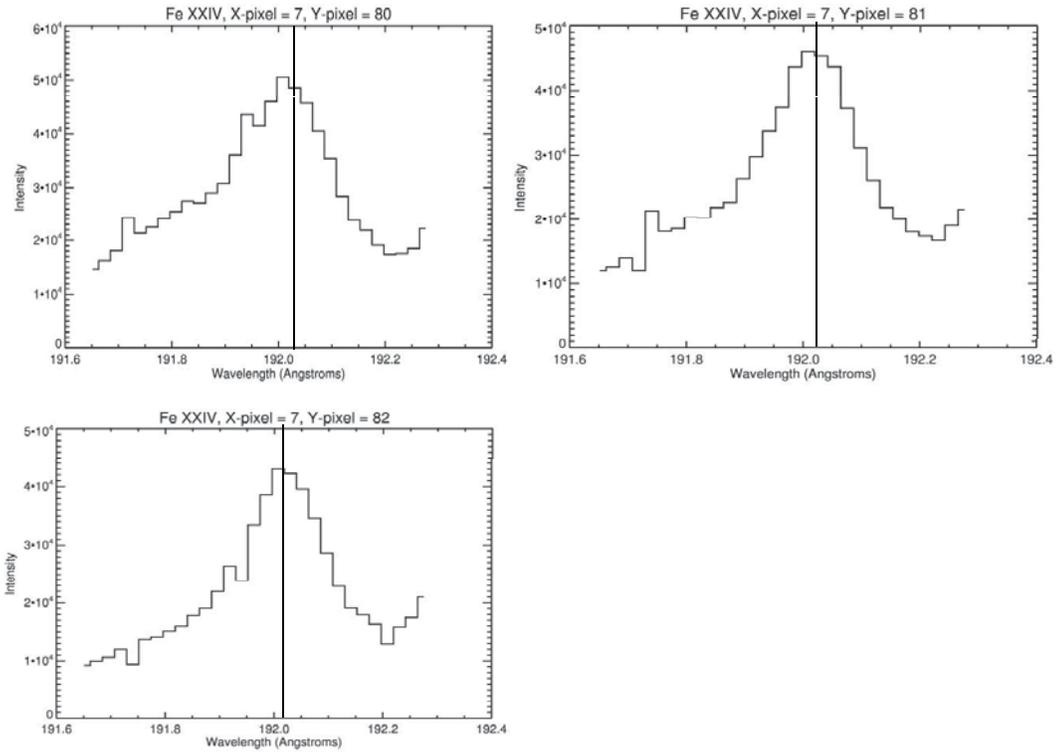

Fig. 18.— Appendix Figure 4: The Fe XXIV line as a function of entire raster Y-pixel for the upflow region marked by the dashed arrow in Figure 4. The Y-pixels in all the Appendix figures equal the Y-pixels in the cutouts of Figure 4 + 50. This is also true for the top panels of Figure 3. The vertical lines in the Appendix figures mark the rest wavelength of the Fe XXIV line.